\pdfoutput=1
\documentclass[3p,times,authoryear]{elsarticle}

\usepackage{amsmath,amssymb}
\usepackage[utf8x]{inputenc}
\usepackage{multirow}
\usepackage{unitsdef}
\usepackage{diagbox}
\usepackage{xcolor}
\usepackage{hyperref}
\hypersetup{
  colorlinks   = true, 
  urlcolor     = blue,
  linkcolor    = blue, 
  citecolor   = red 
}
\newcommand*{\Fig}[1]{Figure~\ref{#1}}
\newcommand*{\Figs}[1]{Figures~\ref{#1}}
\newcommand*{\Table}[1]{Table~\ref{#1}}

\newcommand*{\Eq}[1]{Equation~(\ref{#1})}

\newcommand*{\Sect}[1]{Section~\ref{#1}}

\usepackage{setspace}
\journal{ISPRS}

\biboptions{authoryear}

\begin{document} 

\begin{frontmatter}
\title{Ill-posed Surface Emissivity Retrieval from Multi-Geometry Hyperspectral Images Using a Data- and Theory- driven Hybrid Deep Neural Network}

    \author[addressA,addressB]{Fangcao Xu \corref{CorrespondingAuthor}}  
    \cortext[CorrespondingAuthor]{Corresponding author}
    \ead{xfangcao@psu.edu}
    \author[addressA,addressB,addressC]{Jian Sun}
    \author[addressA,addressB]{Guido Cervone}
    \author[addressD]{Mark Salvador}
    
    \address[addressA]{Department of Geography, The Pennsylvania State University,  University Park, PA, USA}
    \address[addressB]{Institute for Computational and Data Sciences, The Pennsylvania State University,  University Park, PA, USA}
    \address[addressC]{Department of Geosciences, The Pennsylvania State University,  University Park, PA, USA}
    \address[addressD]{Zi Inc, Washington D.C., USA}
		
\begin{abstract}
Atmospheric correction is a fundamental task in remote sensing because observations are taken either of the atmosphere or looking through the atmosphere. Atmospheric correction errors can significantly alter the spectral signature of the observations, and lead to invalid classifications or target detection.  This is even more crucial when working with hyperspectral data, where a precise measurement of spectral properties is required. State-of-the-art physics-based atmospheric correction approaches require extensive prior knowledge about sensor characteristics, collection geometry, and environmental characteristics of the scene being collected. These approaches are computationally expensive, prone to inaccuracy due to lack of sufficient environmental and collection information, and often impossible for real time applications.
	
In this article, a geometry-dependent hybrid neural network is proposed for automatic atmospheric correction using multi-scan hyperspectral data collected from different geometries. Atmospheric correction requires solving the downwelling, upwelling and transmission components of the atmosphere contemporaneously. However, these are dependent on each other. This leads to ill-posed mathematical problem solutions when the training data contain noise and missing observations. The proposed network maintains the physical relationship among different radiative components from the observed at-sensor total radiance. It is equipped with a novel network structure with causality implemented and a loss function, which are trained using two different longwave (7.5 - 13.5$\mu$m) hyperspectral dataset, one simulated using MODTRAN, and the second observed using the Blue Heron instrument in a dedicated field study. The proposed network can characterize the atmosphere without any additional meteorological data.

A grid-search method is also proposed to solve the temperature emissivity separation problem. Results show that the proposed network has the capacity to accurately characterize the atmosphere and estimate target emissivity spectra with a Mean Absolute Error (MAE) under 0.02 for 29 different materials. This solution can lead to accurate atmospheric correction to improve target detection for real time applications.
\end{abstract}
		
\end{frontmatter}

\section{Introduction}
The goal of hyperspectral image analysis is to detect and identify materials based on their spectral signature. In contrast to single- or multi- spectral sensors, hyperspectral sensors measures radiance in hundred of continuous narrow spectral bands where each pixel contains a spectra. With this rich spectral information, it is possible to spectroscopically identify surface materials. 

The ability to leverage surface material separation and identification in remote sensing hyperspectral images is useful for a wide range of scientific fields, such as geologic mapping, urban or forestry monitoring, agricultural crop growth and disease evaluation, emergency response in natural disasters, and weapons of mass destruction detection \citep{ziemann2015hyperspectral, rodrigues2018multi, erkinbaev2019single, shimoni2019hypersectral, ye2019landslide}. Another prominent application is  greenhouse gas detection and global carbon budget modeling \citep{pathakoti2018estimation}.

The major breakthrough in hyperspectral image analysis translates the target detection problem into a classification problem via machine learning to address the large-scale image retrieval problem \citep{gu2019survey}. The abundant spectral-spatial information in multi-scan hyperspectral images enables using fully data-driven methods, such as deep learning neural networks (DNNs). More recently, various deep learning algorithms, especially convolutional neural networks (CNNs), have shown their much stronger feature representation and classification power in the hyperspectral image analysis. Additionally, the end-to-end framework of the deep learning method guarantees an automatic procedure and high efficiency.

One area follows the pathway of Computer Vision and Digital Signal Processing \citep{gao2018hyperspectral, vaddi2020hyperspectral} to train a neural network for pixel classification using its spatial and spectral information. For a large number of labeled training datasets, CNN is most effective and performs well for many applications in remote sensing for the feature extraction and classification by learning high-level semantic features automatically from the whole image to obtain the deep representation. For example, the ImageNet dataset has more than 145 million images were tagged in 20,000 classes \citep{krizhevsky2012imagenet}, which can guarantee enough samples for an end-to-end deep neural network training.

Computer vision algorithms can learn deep representations automatically but still require a great amount of labeled data (e.g., ImageNet) in each material class and their prediction ability is also limited to the pre-defined classes. If the target of interest has never been seen in the training process, the model fails to make predictions. This problem is exacerbated due to publicly available HSI labeled dataset (for training and benchmarking) being very few and also quite outdated \citep{signoroni2019deep}. Another significant problem is that geographic objects in the same class often have a high variety of sizes, colors, and angles in the high resolution hyperspectral images. Alternatively, perceived objects that have similar appearance, such as intensity, color, and orientation, may represent different materials \citep{liang2018material}. The existing computer vision algorithms cannot provide a good solution to these problems because they mostly focus on the colored images and the object spatial information rather than the high-dimensional spectral information.

Although computer vision algorithms can perform high-level perception tasks, such as feature extraction and image recognition, the inner mechanism of these hierarchical information-processing systems is poorly understood in the network generalization \citep{ukita2020causal}. They may be limited in a realistic non-ideal operational use. Most predictive neural network models, especially for data-driven algorithms, cannot answer causal questions because they make predictions using statistics and correlations, but correlation does not imply causation \citep{kleinberg2015guide}. Measures of correlation are symmetrical, because correlation only tells us that there exists a relation between variables. In contrast, causation is usually asymmetrical and therefore gives the directionality of a relation \citep{nauta2019causal}. Having insight into the causal associations in a complex system facilitates the knowledge discovery.

The problem of data-driven approaches has been recognized by many researchers and they started to develop explainable AI \citep{samek2019explainable, wang2019designing}, but many of them employ a mix of causal and human intervened strategies that give inadequate explanations. To implement the causality into the deep learning, a domain-knowledge- or theory- driven approach should be exploited, whereby the causal relationship is given by researchers with theoretical knowledge about how different components are directional connected in the observed data. Then, this causality needs to be coded and implemented into the network structure by breaking down the task into sub-tasks to enable the network to learn the causal reasoning independently while maintaining the inner-relationship at a certain level.

Another area in deep learning is to retrieve the target reflective or emissive spectra from the observed radiance using DNNs for all pixels automatically without labeled data and pre-defined classes. This requires prior knowledge about the spectral characteristics of the desired targets. In these situations, the target spectral characteristics can be defined by a target emissive or reflective spectrum. A radiative transfer model is required to convert the measured at-sensor radiance to the target signal. It describes how the emissive spectrum of a material is sensed by the sensor as radiance, taking consideration of its 3-dimensional surroundings, as well as detailed physical and chemical atmosphere profiles (e.g., gas and aerosols). 

There are several radiative transfer models designed for different imaging systems and spectral-spatial resolutions, such as ACORN \citep{miller2002performance}, HATCH \citep{qu2003high}, FLAASH \citep{vibhute2015hyperspectral}, and MODTRAN \citep{berk2017validation}. The main uncertainty of this conversion via a radiative transfer model comes from sensor calibration errors and noise, atmospheric characterization, compensation for collection geometry distortions, as well as the difficulty associated with the temperature emissivity separation (TES) \citep{pieper2017performance}. This means that to achieve accurate retrieval of atmospheric components from hyperspectral remote sensing data, it is necessary to account for these uncertainties. 

To illustrate this problem, a radiative transfer model is defined as $L(\lambda, \alpha, \beta)$, where $\lambda$ is spectral wavelengths, $\alpha$ represents surface relevant variables (e.g., emissivity, temperature, opacity) and $\beta$ represents solar, atmospheric, and geometric uncertainties (e.g., elevation angle, azimuth, range from the sensor to pixels). A comprehensive $L(\lambda, \alpha, \beta)$ model involves a large number of parameters characterizing the surface, the sensor, the solar radiance, atmospheric, characteristics as well as geometric parameters. The parameter estimation is usually challenging and computationally expensive for many real-world applications.

Many simplifying assumptions or expedient processing steps are made to solve the surface emissive spectra from $L(\lambda, \alpha, \beta)$. First, radiative transfer models simulate the atmosphere as homogeneous plane-parallel layers \citep{lenoble2007successive, efremenko2013, saunders2018update} or using Monte-Carlo simulations \citep{siewert2000Pstar, cornet2010three, russkova2017optimization}, which are however highly computationally expensive and was shown to achieve mixed success. Second, many solutions only focus on specific region of the spectrum, like visible and near-infrared \citep{hakala2018direct}, shortwave \citep{garaba2018sensing}, or longwave infrared \citep{manolakis2019longwave}, because  assumptions can be taken to simplify the problem. Moreover, a single geometric solution is usually applied to every pixel of a spectral image and every image is analyzed individually. This individual image processing is due to limitation in existing solutions, also because collection platforms generally capture data using a nadir looking scan of the target area, and with limited overlaps among scans.

Modeling emissive spectra in the pixel domain from hyperspectral images under different geometries is still a less-explored area due to the complexity of atmospheric correction and surface spatial-spectral and temperature features. This paper proposes to analyze longwave infrared (LWIR) spectra relative to the same surface pixels acquired using multiple scenes collected in rapid sequence and from different angles and ranges. This type of collection assumes agile collection of hyperspectral image such as a gimballed hyperspectral sensor. In this agile collection, altitude, elevation, azimuth, and range to surface pixels of interest vary during collection. New sensors are designed with these capabilities, and overcome the fixed nadir looking geometries limitations of past instruments. These new agile sensors call for the development of new advanced algorithms that can analyze multiple scenes simultaneously and quickly.

There are currently only limited proposed solutions that also take spectra acquired simultaneously from multiple geometries into consideration. \citep{xu2020multiple} proposed an encoder-decoder neural network to retrieve surface emissivity by removing atmospheric effects from LWIR multi-scan hyperspectral data. \citep{sun2021automatic} exploited a time-dependent CNN for reflectivity retrieval in the visible and near infrared hyperspectral images, which outperforms the state-of-the-art methods. However, they are both based on simulated data and have not been applied to the real-world collected data. 

In the real-world data collection process, it is impossible and also not feasible to measure downwelling, upwelling, and atmospheric transmission as well as the temperature and emissivity for every pixel. The missing atmospheric measurements cause an ill-posed problem, meaning that it is not possible to predict each atmospheric component given only the observed at-sensor radiance collection geometry. In other words, multiple solutions for the atmospheric correction can explain the same joint distribution of the observed at-sensor radiance. Besides, the data calibration and the pixel alignment with ground measurements always have a margin of error. To solve this problem, a fully data- and theory- driven network is needed that is more robust on noise and missing data but can better understand the relationship and dependencies of each radiative component in $L(\lambda, \alpha, \beta)$ on various spatial-spectral factors under multi-scan geometries.

A geometry-dependent neural network with the causality implemented into network structure is used in this article to solve every radiative component in $L(\lambda, \alpha, \beta)$ from the observed at-sensor total radiance for the real-world collected LWIR hyperspectral images. The proposed network is partially convolution-based, but involves two geometric factors, i.e., collection range and angle, using the fully connected layers. The geometry-dependency of such a neural network is expected to capture the physical characteristics of atmospheric absorption and scattering under different geometries, which can improve the accuracy of estimating the target emissivity spectra when the data contains noise or missing information. 

This paper is organized as follows: first, the real-world collected LWIR images as well as the simulated data are described; second, the theoretical radiative transfer model and method to separate the surface temperature and emissivity are introduced; finally, the architectural design of the geometry-dependent neural network as well as the training process are discussed. The performance of our designed neural network is then evaluated by comparing the ground-truth radiative components in $L(\lambda, \alpha, \beta)$ and the predicted results. Surface temperature and emissivity are also retrieved as our final step to evaluate the network's ability to map the observed at-sensor radiance spectra to the surface emissive spectra. 

\section{Data Description}
Data collection field experiments are costly and time consuming, and require a considerable investment in planning, acquiring and analyzing results. They generally involve making aerial and ground truth measurements at the same time. Besides, collected data, no matter how controlled the experiment is, generally always contain noise and missing observations. Hyperspectral data collections in particulars are expensive, and even more so in the longwave part of the spectrum due to the cost of the instruments.

In this research, two different datasets are employed to train the network: 1) $1.4\ TB$ real-world collected Blue Heron (BH) longwave hyperspectral images, and 2) $196\ G$ MODerate spectral resolution TRANsmittance (MODTRAN) simulated data for 29 different materials (e.g., Aluminum, Grass, Cropland etc.) at different elevation angles and ranges. 

Simulated hyperspectral data are included in this study because at the time of the real world data collection, a precise characterization of the atmosphere, and even more so for each pixel, could not be completed. While downwelling and upwelling can be collected during field experiments, transmission is highly complex and requires atmospheric soundings using balloons or lasers. The BH data only measured the total radiance received at the sensor and the geometric parameters for each pixel. Thus, solving more than one atmospheric components from only one observed at-sensor variable under different geometries is an ill-posed problem. The data-driven deep learning algorithms are extremely limited in this non-ideal operational use. A hybrid network structure is proposed to exploit the simulated data which have ground-truth atmospheric radiance measurements and target information. This allows the network to learn the missing information in the real-world collected data and can greatly improve the efficiency and lower the cost of the data collection and analysis process.

\subsection{NittanyRadiance-2019 Field Experiment}
The NittanyRadiance-2019 field data collection experiment was performed using the Harris Blue Heron longwave hyperspectral imaging spectrometer over the Pennsylvania State University campus on April 18, 2019. Blue Heron is a high spatial resolution gimballed hyperspectral sensor. The sensor has two LWIR focal plane arrays each collecting 256 spectral bands within the 7.5 to 13.5 $\mu m$ spectral range. It also collects visible data using a wide angle optical sensor at 30 Hz and with a $0.5\ m$ spatial resolution at 10,000 feet.

The Blue Heron sensor was flown on a Cessna Grand Caravan which flew from Rochester, NY, to State College, PA on the day of the experiment. The data collection started around 1:00 pm (local time) and lasted for over two hours. The plane was at a nominal altitude of about 10,000 feet AGL. The weather was generally sunny with occasional thin cirrus clouds.

\begin{figure}[ht]
	\centering
	\includegraphics[width=0.9\textwidth]{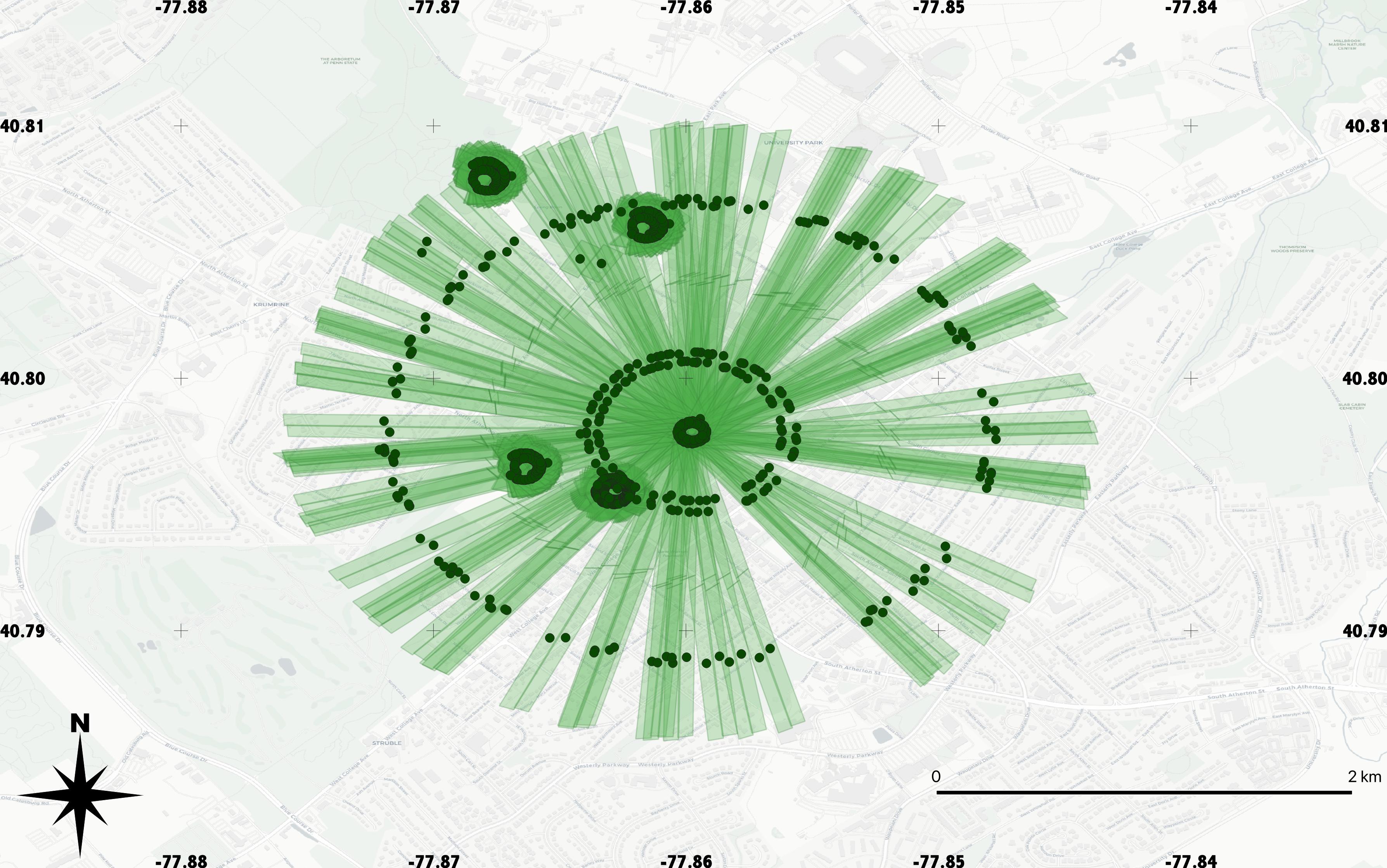}
	\caption{Blue Heron Data Collection Orbit and Collected Images}
	\label{fig:BHcollect}
\end{figure}

\Fig{fig:BHcollect} shows the footprints for the 1,436 longwave infrared hyperspectral scenes collected along the plane orbit area. The sensor collected long-narrow images over the whole campus area, and many high resolution scenes for four specific areas where ground targets were installed.

The number of pixels in each collected hyperspectral image varies from 7,998 to 464,400, with a mean value of 166,157. Each pixel has the at-sensor total radiance collected at 256 wavelength bands from $7.5\ \mu m$ to $13.5\ \mu m$ as well as two coordinates and one DEM value. This means that, on average, each image contains $166,157*(256+3)$ spectral and spatial measurements. In addition to sensor measurements, metadata include telemetry of the BH sensor, approximate geo-locations for the sensor and some of the target pixels. Unfortunately, the metadata do not include the banking angle of the plane at the time of the collection, but only the angle of the instrument with respect to the axis of the plane.  This angle was estimated through simple trigonometry, and used to correct the collection geometry angle.

The data were calibrated by the Harris Corporation according to sensor specifications, and the geometric relationship between the BH sensor and each pixel in the collected hyperspectral scenes were computed and expressed by the range, angle parameters. \Table{tb:parasBH} lists all parameters and their values observed in BH collected images.
\begin{table}
    \begin{center}
    \caption{Parameters in Blue Heron Collections}\label{tb:parasBH}
    \begin{tabular}{ l|l }
    \hline
    \bf Parameters & \bf Values\\
    \hline
    Wavelength $\lambda$ & $[7.5 - 13.5]$ by every $0.0234$ ($\mu m$)\\
    Day of the year & $108$\\
    Time of the day & $[13:00 - 15:00]$ (EST)\\
    Elevation Angle $\theta$ & $[30 - 60]$ $(^\circ)$\\
    Ranges & $[3600 - 6150]$ $(m)$\\
    \hline
    \end{tabular}
    \end{center}
\end{table}

\begin{figure}[ht]
	\centering
	\includegraphics[width=\textwidth]{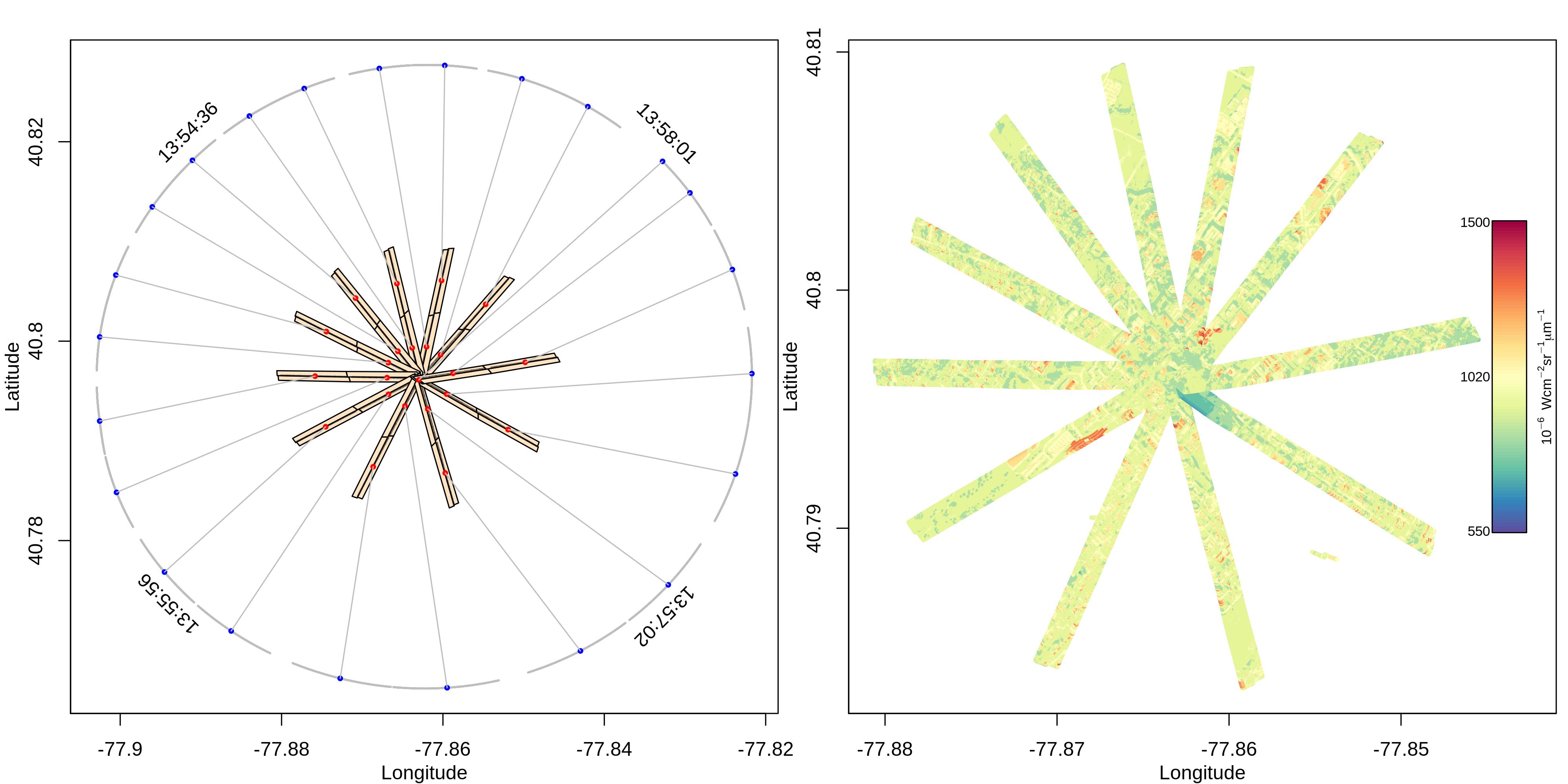}
	\caption{Blue Heron Hyperspectral Images and Geometric Relationship}
	\label{fig:BHimages}
\end{figure}

\Fig{fig:BHimages} shows 46 hyperspectral images taken during one loop of the plane. The blue and red points in the left plot are the centers of the sensor paths and the centers of corresponding pairs of images collected along sensor paths. There are always two images collected simultaneously because the BH sensor has two focal plane arrays. One orbit of collection occurs in approximate five minutes intervals, during which time the atmosphere is invariant. The right plot in \Fig{fig:BHimages} is the spectral radiance map at $10.1\ \mu m$ in the unit of micro flick ($10^{-6}*[W cm^{-2} sr^{-1} um^{-1}]$).

\begin{figure}[ht]
	\centering
	\includegraphics[width=\textwidth]{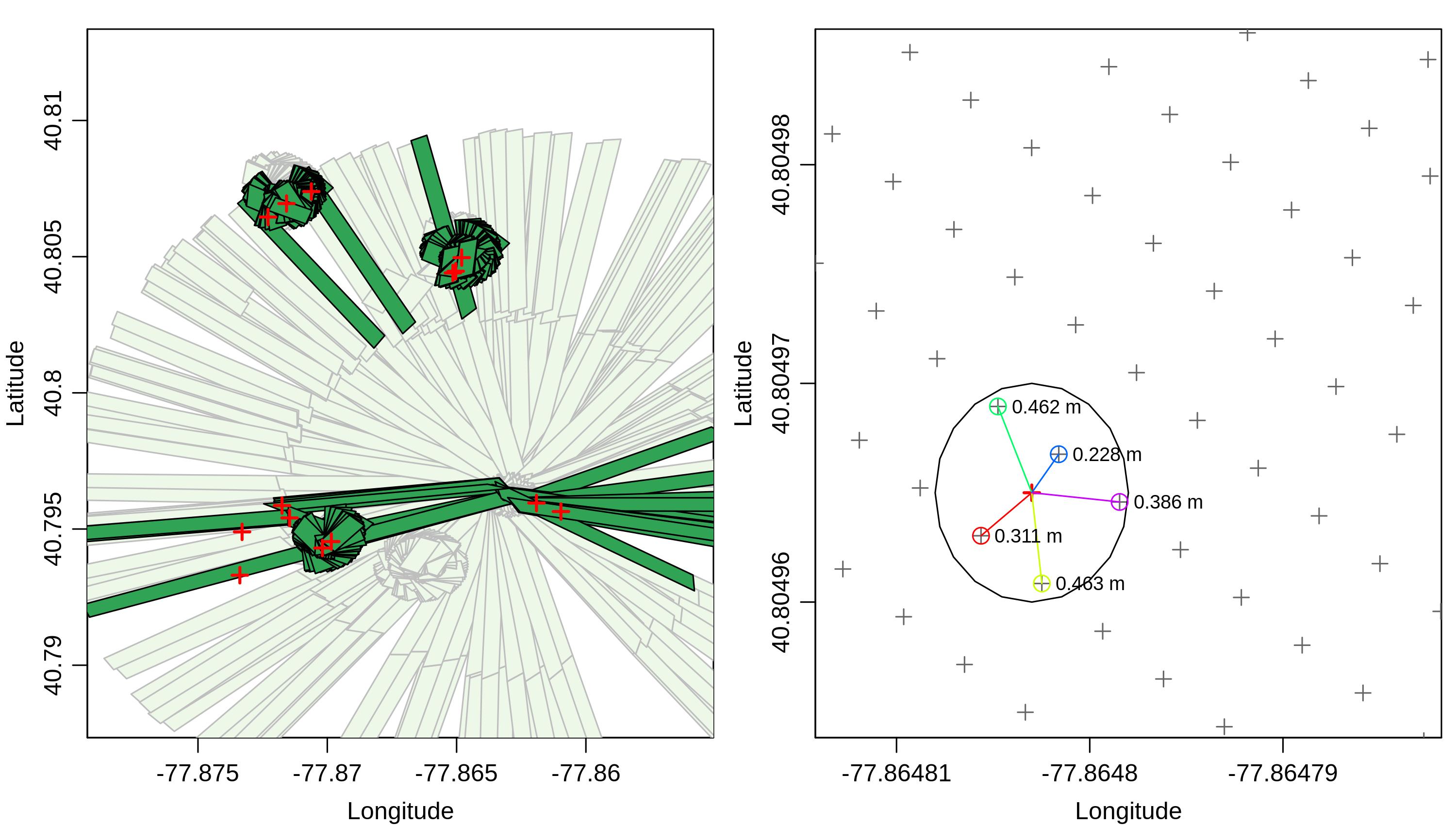}
	\caption{Extracted Pixels for given Target in Blue Heron Hyperspectral Images}
	\label{fig:pixels}
\end{figure}

\Fig{fig:pixels} shows the process to extract pixels with known spectral emissivity from BH images as the training dataset. Red points and green polygons in the left plot are 14 locations selected for the grass pixels and their corresponding hyperspectral images intersecting with them. Given each location, a $0.5\ m$ searching radius is used to extract nearby pixels from each overlaid image as the labelled data for grass, shown in the right plot. 

The aggregated information from different hyperspectral images as grass pixels for training the proposed network is listed in \Table{tb:BHdata}, including image ID, image side (e.g., side 1=left or 2=right), pixel longitude, latitude and altitude above ground, the calculated range and elevation angle between the pixel and the sensor, and the at-sensor total radiance over 256 wavelength bands.  
\begin{table*}[!ht]
    \includegraphics[width=\textwidth]{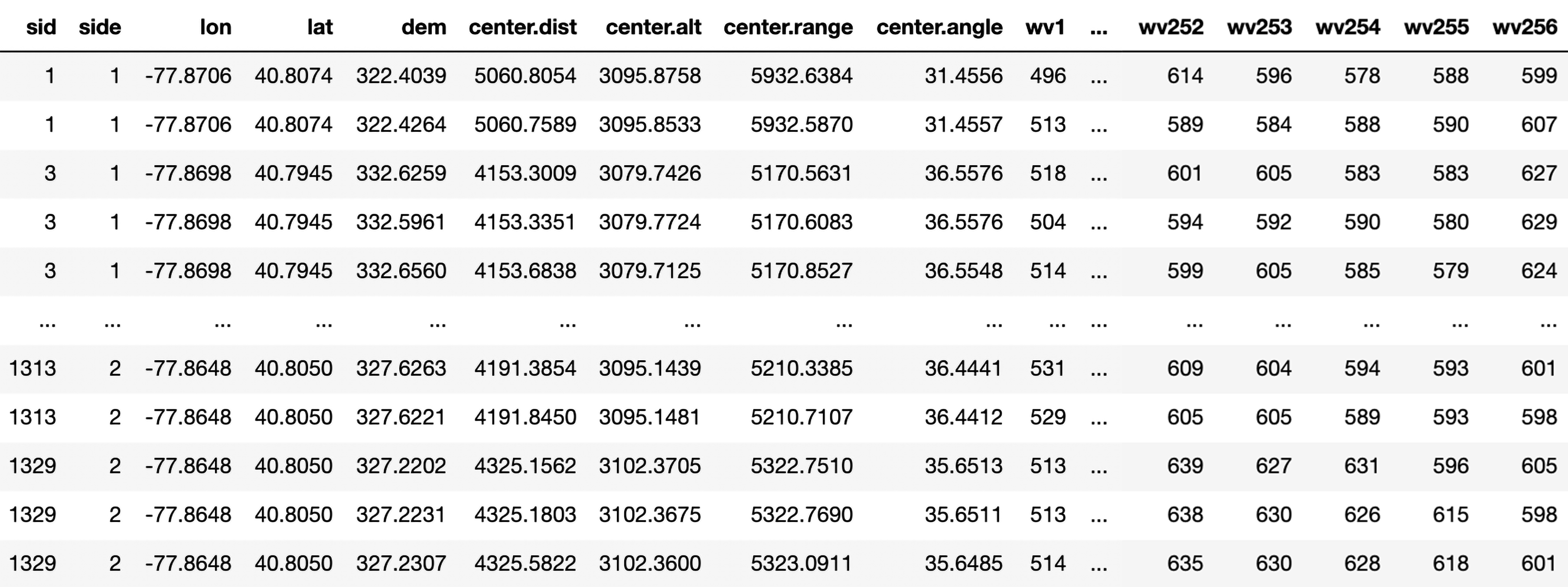}
    \caption{Data Table of Extracted Pixels from Multiple Hyperspectral Images}
    \label{tb:BHdata}
\end{table*}

\subsection{MODTRAN Simulated Data}
MODTRAN is a physics-based radiative transfer model co-developed by Spectral Sciences, Inc. and the Air Force Research Laboratory (AFRL) to simulate optical measurements through the atmosphere. It computes line-of-sight (LOS) atmospheric spectral transmission and radiances from the ultraviolet (UV) through long wavelength infrared (IR) spectral range \citep{berk2014modtran}.

MODTRAN was used to simulate different atmospheric scenarios to help solve the ill-posed atmospheric characterization of BH collected data. MODTRAN  provides accurate values for different radiative components from multiple geometries if detailed atmospheric parameters can be provided. The output of MODTRAN is then used to train the deep learning network for finding the local optima which reflect the real-world physical relationships among the radiative components. In other words, MODTRAN is used to constrain the search space spanned by the deep learning solution.

The simulations were performed assuming an instrument with the same spectral resolution of the BH sensor, namely of 23.4 $nm$ from 7.5 to 13.5 $\mu m$. Each simulation assumed a different angle and range for 29 different known built-in materials.  Specifically, simulations were made for 31 elevation angles from $30^\circ$ to $60^\circ$ with a resolution of $1^\circ$, and 36 ranges from $3000\ m$ to $6500\ m$ with a resolution of $100\ m$  (\Table{tb:parasMODTRAN}).

\begin{table*}
    \begin{center}
    \caption{Parameters in MODTRAN Simulation}\label{tb:parasMODTRAN}
    \begin{tabular}{ l|l }
    \hline
    \bf Parameters & \bf Values\\
    \hline
    Wavelength $\lambda$ & $[7.5 - 13.5]$ by every $0.0234$ ($\mu m$)\\
    Day of the year & $108$\\
    Time of the day & $[10:00,\ 12:00,\ 14:00,\ 16:00]$ (EST)\\
    Targets & 29 Built-in Materials\\
    Target Temperature $T$ & $[295,\ 300,\ 305,\ 310,\ 315,\ 320]$ $(Kelvins)$\\
    Elevation Angle $\theta$ & $[30 - 60]$ by every $1$ $(^\circ)$\\
    Ranges & $[3000 - 6500]$ by every $100$ $(m)$\\\hline
    \end{tabular}
    \end{center}
\end{table*}

\begin{figure}[!ht]
	\centering
	\includegraphics[width=\textwidth]{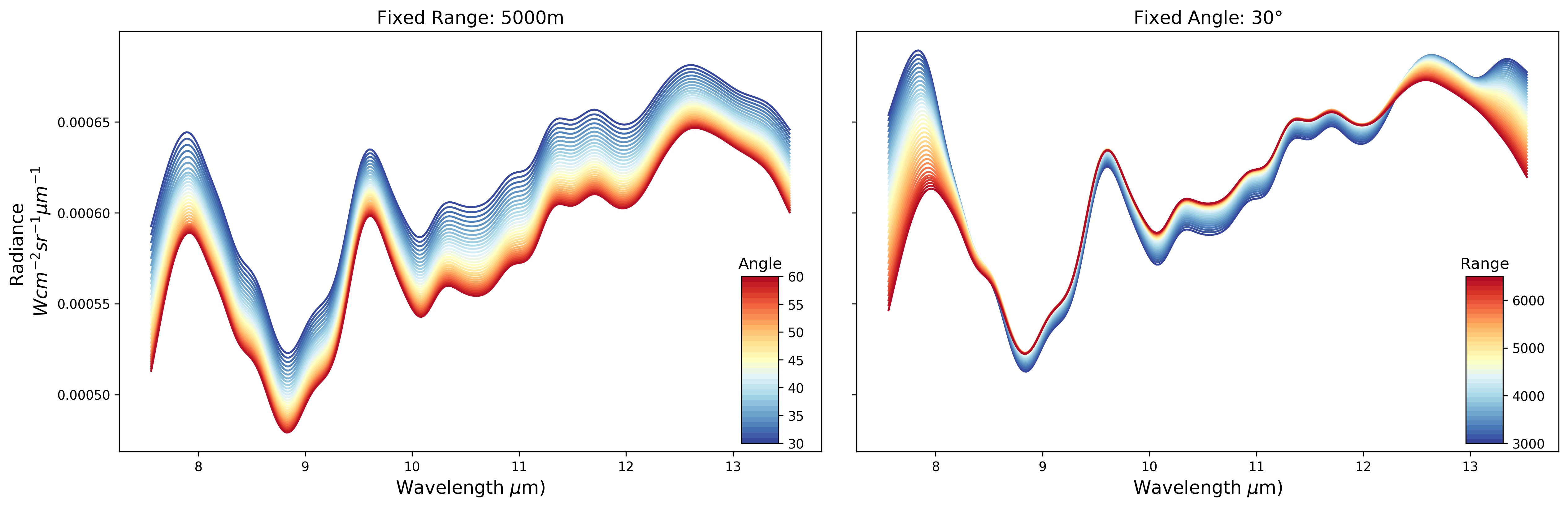}
	\caption{MODTRAN simulated at-sensor total radiance with various angles and ranges for Aluminum}
	\label{fig:MODTRANgeo}
\end{figure}

\Fig{fig:MODTRANgeo} shows the simulated at-sensor total radiance under different geometries. The left plot is with different elevation angles but a fixed range while the right plot is with different ranges but a fixed elevation angle. 

For a fixed range, the at-sensor total radiance increases when elevation angle decreases because the line-of-sight (LOS) path  goes through warmer and denser air resulting in a higher atmospheric thermal emission and scattering. For a fixed elevation angle, the at-sensor total radiance is inverted twice across the different ranges as the wavelength increases. Even though atmospheric thermal emission is higher over the longer range, the transmission is lower because the energy is absorbed or scatter by the atmosphere over this larger range. At wavelengths where the atmospheric absorption is stronger than the emission, the at-sensor total radiance decreases with an increase in range. 

\section{Methods}
\subsection{Multi-Geometry Radiative Transfer Model}
The parameter estimation of radiative transfer model in the LWIR spectrum is more complicated compared to visible or shortwave infrared spectrum because the surface emits and reflects radiance at the same time. The measured at-sensor radiance is a function of the target emissivity and temperature of the surface, the reflected downwelling radiance, the upwelling radiance, and the atmospheric transmission. Besides, the hyperspectral remote sensing data are also highly affected by the sensor geometries during collections and by the absorption and scattering of the atmosphere, including gases, aerosols, water vapor, and clouds. The physical transportation of radiative energy can be described by the radiative transfer equation, and its mathematical expression \citep{schott2007remote} is listed below:
\begin{equation}\label{equ:RTE}
\begin{aligned}
    L_\lambda & = [\frac{cos\sigma}{\pi}E_{s\lambda}\tau_1(\lambda) + F(L_{ds\lambda} + L_{d\epsilon\lambda}) \\
    & + (1-F)(L_{bs\lambda}+ L_{b\epsilon\lambda})]r(\lambda)\tau_2(\lambda)  \\
    & + L_{T\lambda}\varepsilon(\lambda)\tau_2(\lambda) + L_{us\lambda} + L_{u\epsilon\lambda}\\
\end{aligned}
\end{equation}
where $T$ is the target is temperature in Kelvin, $s$ represents that the component is solar-related, $\epsilon$ denotes the self-emitted thermal component, $d$ ($u$) is the downwelling (upwelling) component, and $b$ is the background involved component. $F$ is the fraction of hemisphere that is not obscured by background objects, and is also known as the shape factor. $\lambda$ denotes the spectral wavelength. The physical meaning of each RTE component is listed in \Table{tb:paras_RTE}.

\begin{table*}[!ht]
	\centering
	\caption{Physical meaning of each component in RTE} 
	\label{tb:paras_RTE}
	\begin{tabular}{ l|l|l}
		\hline
		\bf Components & \bf Units & \bf Physical meaning \\
		\hline
		$E_{s\lambda}$  & $W/m^2$   & Extraterrestrial solar irradiance \\  
		$\sigma$        & -       & Incident angle of solar irradiance \\
		$r(\lambda)$    & -       &  Spectral reflectivity of the target \\
		$\varepsilon(\lambda)$  & -    & Spectral emissivity of the target \\
		$\tau_1(\lambda)$ 	    & - 	 & Atmospheric transmission on the sun-target path \\
		$\tau_2(\lambda)$ 	    & - 	 & Atmospheric transmission on the target-sensor path\\
		$L_{T\lambda}$ 			               & $Wcm^{-2}sr^{-1}\micrometer^{-1}$ & Spectral radiance of a blackbody at temperature $T$\\
		$L_{ds\lambda} + L_{d\epsilon\lambda}$ & $Wcm^{-2}sr^{-1}\micrometer^{-1}$ & Sum of solar and atmospheric downwelling spectral radiance \\
		$L_{bs\lambda} + L_{b\epsilon\lambda}$ & $Wcm^{-2}sr^{-1}\micrometer^{-1}$ & Sum of background reflected and self-emitted spectral radiance \\
		$L_{us\lambda} +L_{u\epsilon\lambda}$  & $Wcm^{-2}sr^{-1}\micrometer^{-1}$ & Sum of solar and atmospheric upwelling spectral radiance \\
		\hline
	\end{tabular}
\end{table*}

For simplicity, two assumptions are made in the following contents: 1) all targets of interest are in open area, where the shape factor $F = 1$, and 2) target surfaces are opaque. Conservation of energy indicates that all incident flux must be either transmitted, reflected, or absorbed, i.e., $\tau(\lambda) + r(\lambda) + \alpha(\lambda) =1$. For an opaque surface, whose transmittance $\tau(\lambda)=0$ and absorptance equals to the emissivity $\alpha(\lambda) = \varepsilon(\lambda)$, thus the emissivity can be calculated as: $\varepsilon(\lambda)=1-r(\lambda)$. 

Based on the sensitivity study done by \citep{xu2020multiple}, the solar components are at least 3 orders of magnitude smaller than thermal signatures in the longwave infrared spectrum, which is negligible in this research. Thus \Eq{equ:RTE} can be reformulated as:
\begin{equation} \label{equ:simpRTE}
\begin{aligned}
    L_\lambda & =[(1-\varepsilon(\lambda))L_{down} +\varepsilon(\lambda)L_{T\lambda}]\tau(\lambda) + L_{up}  \\
    &= (1-\varepsilon(\lambda))L_{down}\tau(\lambda) + L_{emit} + L_{up}
\end{aligned}
\end{equation}
where, $L_{down}=L_{d\epsilon\lambda}$ describes the downwelling radiance reaching at the target and $L_{emit}=\varepsilon(\lambda)L_{T\lambda}\tau(\lambda)$ describes the self-emitted radiance of the target reaching at the sensor; $\tau(\lambda)$ is the atmospheric transmission from the target to the sensor; $L_{up} =L_{u\epsilon\lambda}$ describes at-sensor upwelling radiance.

$L_{T\lambda}$ is referred as Planck Equation, which is the spectral radiation directly emitted by a blackbody at temperature $T$:
\begin{equation}\label{equ:Planck}
    L_{T\lambda} = 2hc^2\lambda^{-5}(e^{\frac{hc}{\lambda kT}}-1)^{-1}
\end{equation}
where $h$ is the Planck's constant ($6.6256×10^{−34}\ Js$), $c$ is the speed of light ($2.998×10^8\ ms^{−1}$), and $k$ is the Boltzmann gas constant ($1.381×10^{−23}\ J K^{−1}$). Given the sensor spectral bands, the only unknown variable to decide the $L_{T\lambda}$ is the target temperature $T$.

To characterize the atmosphere, a desired network should be capable of predicting $L_{down}$, $L_{up}$, and $\tau(\lambda)$, which are independent of surface materials. The downwelling radiance $L_{down}$ is the integrated atmospheric emittance over the hemisphere reaching at the surface. It is geometry-independent for all pixels collected in a short time window, assuming an open area under invariant atmosphere. Thus, it can be written as a function of wavelength. $L_{up}$ and $\tau(\lambda)$ are variables measured from the surface to the sensor after considering atmospheric attenuation under different geometries so they are dependent on the sensor-target geometry (i.e., range, angle). Surface emittance reaching at the sensor, $L_{emit}$ is a function of surface emissivity $\varepsilon(\lambda)$, temperature $T$ and the predictable $\tau(\lambda))$.

\begin{equation}\label{equ:DL}
\begin{split}
    L_{down} &= \Omega_1(wavelength) \\
    L_{up}, \tau(\lambda) &= \Omega_2(wavelength,\ range,\ angle)\\
    L_{emit} &= \Omega_3(T, \varepsilon(\lambda), \tau(\lambda)) \\
\end{split}
\end{equation} 

Once the atmospheric components (i.e., $L_{down},\ L_{up},\ \tau(\lambda)$) have been accurately quantified, it is possible to further estimate the two remaining unknown target-dependent variables (emissivity $\varepsilon(\lambda)$ and temperature $T$).

\subsection{Deep Learning Solution for RTE}
The proposed network introduces this causality into the network structure to understand how different radiative components and geometric factors rely on each other and contribute together towards the at-sensor total radiance. The network fully exploits multi-scan geometric information, the physical relationship among different radiative components in \Eq{equ:simpRTE}, and the observed at-sensor total radiance to achieve atmospheric characterization and target detection without any meteorological data. To be more specific, the pixel-wise available information in BH hyperspectral images only includes: 1) the pixel and sensor's temporal coordinates, 2) the observed range and angle between the pixel and the sensor, and 3) total radiance received at the sensor for the pixel.

\subsubsection{Hybrid Network Structure}
\begin{figure}[ht]
	\centering
	\includegraphics[width=\textwidth]{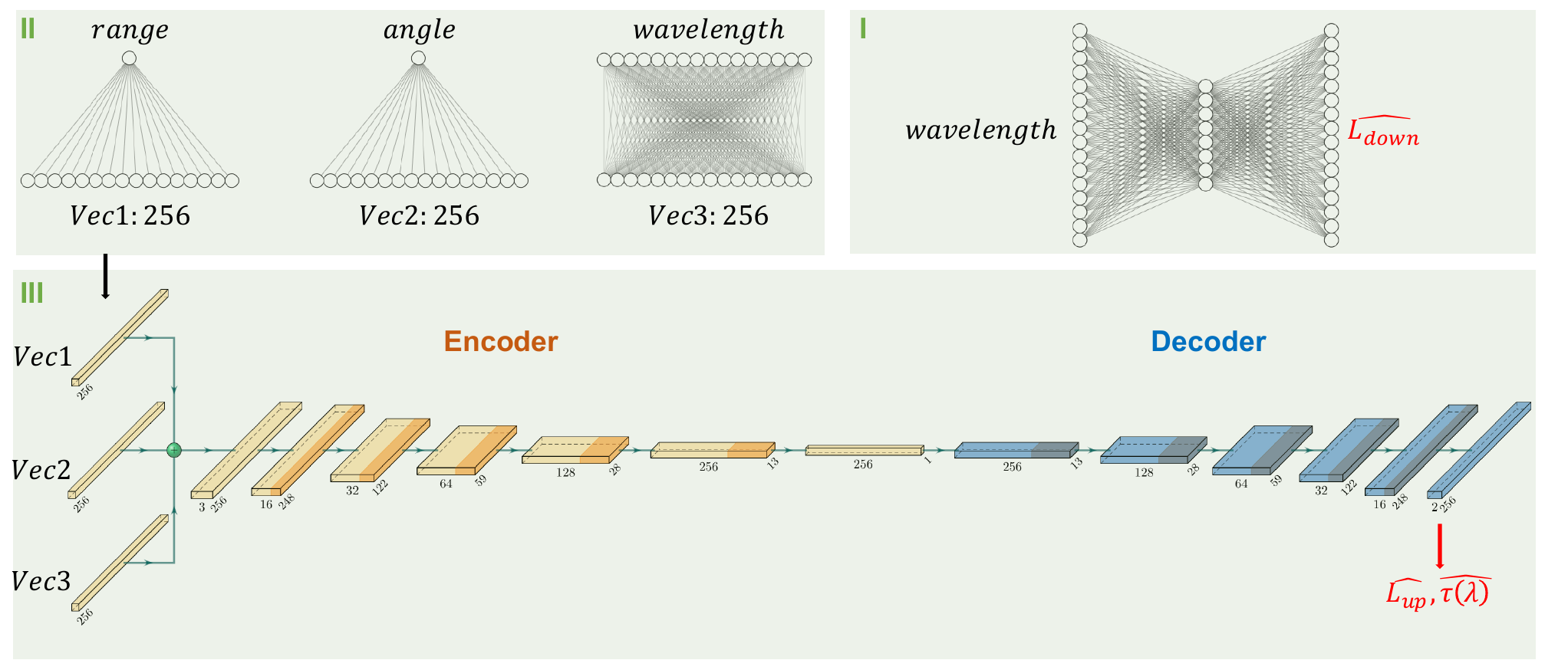}
	\caption{The architecture of the proposed geometry-dependent neural network}
	\label{fig:architecture}
\end{figure}

\Fig{fig:architecture} shows our proposed hybrid network including three primary blocks, 1) a two-layer fully connected network (I), followed by a Sigmoid activation layer at the end to predict the normalized downwelling radiance, 2) three single-layer fully connected networks (II), each followed by a Leaky Rectified Linear Unit (LeakyReLU) activation layer, to transform the range, angle and wavelength into the feature vectors with the same length of 256 neurons, and 3) a convolutional encoder and decoder (III) to take the blended 3x256 feature vectors generated by (II) to predict the normalized upwelling radiance $L_{up}$ and unnormalized transmission $\tau(\lambda)$.

Thus, the network requires three inputs for each pixel, including the wavelength, range and angle. The fully-connected networks (II) directly take these three inputs to generate feature vectors for the encoder-decoder framework, which are key factors to learn the atmospheric characteristics under multiple geometries. The downwelling radiance reaching at the target is the integration of all opening atmosphere self-emitted radiance, for which the range and angle are not the causes. Thus, downwelling radiance is predicted by a distinct network (I), which is constrained by \Eq{equ:simpRTE}. The variation of geometric factors and the wavelengths causes the upwelling radiance and atmospheric transmission vary. The branched architecture of proposed network allows that $L_{up}$ and $\tau(\lambda)$ can be learned separately from $L_{down}$.

This network requires data collected at multi-geometries to learn the geometry-independent $L_{down}$ and geometry-dependent $\tau(\lambda)$ and $L_{up}$. If geometries are same for all collected ground pixels, this proposed hybrid network would be same as other data-driven deep neural networks without the causality implemented. This is because, even the $range$ and $angle$ are the causes for the variation of $\tau(\lambda)$ and $L_{up}$, the observed data only have one geometry. Thus, the network cannot learn the geometry dependencies of different atmospheric components from the observed data.

The at-sensor total radiance $\widehat{L_\lambda}$ is calculated based on the predicted $L_{up}$, $\tau(\lambda)$ and $L_{down}$ via the radiative transfer equation in \Eq{equ:simpRTE}. The network is trained to minimize the error between the calculated $\widehat{L_\lambda}$ and observed $L_\lambda$ for the real-world collected data. Thus, the physical relationship of $L_{down}$, $L_{up}$ and $\tau(\lambda)$ is maintained via the radiative transfer equation towards the $\widehat{L_\lambda}$ under different geometries through the training process.

The encoder in this architecture is designed to remove repetitive information from the multi-scan hyperspectral data by analyzing the invariance of the surface emissivity under different elevation angles and ranges. To preserve the detailed information while reducing spatial and spectral dimensionality of the inputs, instead of using the max pooling layer, strided convolutions are adopted to compress the input into a latent space vector. There are six convolution blocks in the encoder, in which each block contains a strided convolutional layer, a batch normalization layer, and a LeakyReLU activation layer. 

Six transposed convolutional layers in the decoder are used as upsampling strategy, which are approximately reversed processes of convolutional operations in the encoder. A combination of a batch normalization layer and a LeakyReLU transformation are implemented after each transposed convolution layer, except for the last one which is followed by a Sigmoid activation to ensure the predicted normalized upwelling and unnormalized transmission between 0 and 1. 

This proposed network structure can force the model to maintain the physical relationship among different radiative components under multi-scan geometries with the causality implemented. However, even with such a physics-informed framework, characterizing the atmosphere still remains to be an ill-posed problem, i.e., more than one optimal solution of $L_{down}$, $L_{up}$, and $\tau(\lambda)$ may exist to satisfy the \Eq{equ:simpRTE}, for all observations. Thus, training the network with the physics-law (Equation 2) only may lead to a false projection between the input and these radiative components, further improperly characterizing the atmosphere. These uncertainties can either be mitigated by adding additional empirical constraints or by providing the ground-truth atmospheric values during the training process.

In the following experiments, to enforce the network learning the correct projection between the input and radiative components, the MODTRAN simulated dataset is supplemented to train the network with BH collected images. At every epoch, the network is firstly trained on MODTRAN simulated data with the ground-truth atmospheric radiative components and \Eq{equ:simpRTE}, which prevents the network from converging into a false and local optimum during the training process. Then the network is refined using the BH data with \Eq{equ:simpRTE} which ensures the network's capacity of capturing real-world patterns.

\subsubsection{Network Parameters}
All MODTRAN simulated data for 29 different materials and BH extracted pixels for grass are split into $80\%$ as training and $20\%$ as testing. The $80\%$ training dataset is further split into $90\%$ as training and $10\%$ as validation. To balance the network learning ability on MODTRAN and BH dataset which are at different scales, the training dataset in each epoch is as follows: 1) randomly selected data points from the MODTRAN training dataset which are two times the amount of BH training dataset, and 2) all the BH training dataset. 

The network is trained on the NVIDIA Quadro P5000 GPU with four workers, using PyTorch. The detailed parameters, including the size of convolutional filters, stride parameters, and padding strategy, for each convolutional layer are reported in \Table{tb:ParasNetwork}:
\begin{table*}[!htbp]
	\centering
	\caption{Detailed parameters of the proposed geometry-dependent CNN shown in \Fig{fig:architecture}}
	\label{tb:ParasNetwork}
	\begin{tabular}{clccccc}
		\hline
		\multicolumn{2}{c|}{Operation Layer} & 
		\begin{tabular}[c]{@{}c@{}}Number \\ of Filers\end{tabular} & \begin{tabular}[c]{@{}c@{}}Size of \\ Each Filter\end{tabular} & Stride & Padding & \begin{tabular}[c]{@{}c@{}}Output Size\\ (seq x c x l)\end{tabular} \\\hline
		\multicolumn{1}{c|}{\multirow{4}{*}{\begin{tabular}[c]{@{}c@{}}Fully\\ Connected\\ Layers\end{tabular}}} & \multicolumn{1}{l|}{FC1\_wavelength} & - & - & - & - & 256 \\\cline{2-7}
		\multicolumn{1}{c|}{} & \multicolumn{1}{l|}{FC2\_range} & - & - & - & - & 256 \\\cline{2-7} 
		\multicolumn{1}{c|}{} & \multicolumn{1}{l|}{FC2\_angle} & - & - & -  & - & 256 \\\cline{2-7} 
		\multicolumn{1}{c|}{} & \multicolumn{1}{l|}{FC2\_wavelength} & - & - & - & - & 256 \\\hline
		\multicolumn{1}{c|}{Input} & \multicolumn{1}{l|}{Latent Vectors} & - & -  & - & - & (seq, 3, 256) \\ \hline
		\multicolumn{1}{c|}{\multirow{4}{*}{Encoder}} & 
		\multicolumn{1}{l|}{Conv\_layer\_1} & 16  & (11,)  & (1,) & (1,) & (seq, 16, 248) \\ \cline{2-7} 
		\multicolumn{1}{c|}{} & \multicolumn{1}{l|}{Conv\_layer\_2} & 32  & (7,) & (2,) & (1,) & (seq, 32, 122) \\\cline{2-7} 
		\multicolumn{1}{c|}{} & \multicolumn{1}{l|}{Conv\_layer\_3} & 64  & (7,) & (2,) & (1,) & (seq, 64, 59) \\\cline{2-7}
		\multicolumn{1}{c|}{} & \multicolumn{1}{l|}{Conv\_layer\_4} & 128 & (7,) & (2,) & (1,) & (seq, 128, 28) \\\cline{2-7}
		\multicolumn{1}{c|}{} & \multicolumn{1}{l|}{Conv\_layer\_5} & 256 & (5,) & (2,) & (1,) & (seq, 256, 13) \\\cline{2-7}
		\multicolumn{1}{c|}{} & \multicolumn{1}{l|}{Conv\_layer\_6} & 256 & (13,) & (1,) & (0,) & (seq, 256, 1) \\\hline
		\multicolumn{1}{c|}{\multirow{4}{*}{Decoder}} & 
		\multicolumn{1}{l|}{Conv\_trans\_1} & 256 & (13,) & (1,) & (0,) & (seq, 256, 13) \\\cline{2-7} 
		\multicolumn{1}{c|}{} & \multicolumn{1}{l|}{Conv\_trans\_2} & 128 & (6,) & (2,) & (1,)  & (seq, 128, 28) \\\cline{2-7} 
		\multicolumn{1}{c|}{} & \multicolumn{1}{l|}{Conv\_trans\_3} & 64 & (7,) & (2,) & (1,)  & (seq, 64, 59) \\\cline{2-7} 
		\multicolumn{1}{c|}{} & \multicolumn{1}{l|}{Conv\_trans\_4} & 32 & (8,) & (2,) & (1,)  & (seq, 32, 122)  \\\cline{2-7}
		\multicolumn{1}{c|}{} & \multicolumn{1}{l|}{Conv\_trans\_5} & 16 & (7,) & (2,) & (1,)  & (seq, 16, 247)  \\\cline{2-7}
		\multicolumn{1}{c|}{} & \multicolumn{1}{l|}{Conv\_trans\_6} & 2 & (12,) & (2,) & (1,) & (seq, 2, 256)  \\\hline
	\end{tabular}
\end{table*}

\subsubsection{Loss Function}
The MODTRAN simulated dataset has ground-truth measurements of all radiative components, including atmospheric radiative components ($L_{down}$, $L_{up}$, $\tau(\lambda)$) and the at-sensor target emittance and total radiance ($L_{emit}$, $L_\lambda$), while the BH dataset only has the observed $L_\lambda$. Thus, two loss functions are defined to guide the training process:

\begin{equation}\label{equ:lossfunc}
\begin{gathered}
    \widehat{L_{emit}}  = L_T\varepsilon(\lambda)\widehat{\tau(\lambda)}\\
    \widehat{L_\lambda} = (1-\varepsilon(\lambda))\widehat{L_{down}}\widehat{\tau(\lambda)} + \widehat{L_{emit}} + \widehat{L_{up}}\\\\
    \begin{aligned}
        loss1 = &[(L_\lambda - \widehat{L_\lambda})^2 + (L_{down} - \widehat{L_{down}})^2 +(L_{up} - \widehat{L_{up}})^2 \\
        & + (L_{emit} - \widehat{L_{emit}})^2 + (\tau(\lambda) - \widehat{\tau(\lambda)}^2]/5 \\
        loss2 = &(L_\lambda - \widehat{L_\lambda})^2 
    \end{aligned}
\end{gathered}
\end{equation}
where all predicted or derived components are denoted by $\ \widehat{}\ $. The network is trained on surfaces with known temperatures $T$ and emissivity $\varepsilon(\lambda)$, thus the $\widehat{L_{emit}}$ and $\widehat{L_\lambda}$ can also be accurately calculated with predicted and known variables. When the data point is MODTRAN simulated, the $loss1$ function is used to calculate the model error. Otherwise, the $loss2$ function is used for the BH data where only the at-sensor total radiance $L_\lambda$ is acquired. 

\begin{figure}[ht]
	\centering
	\includegraphics[width=\textwidth]{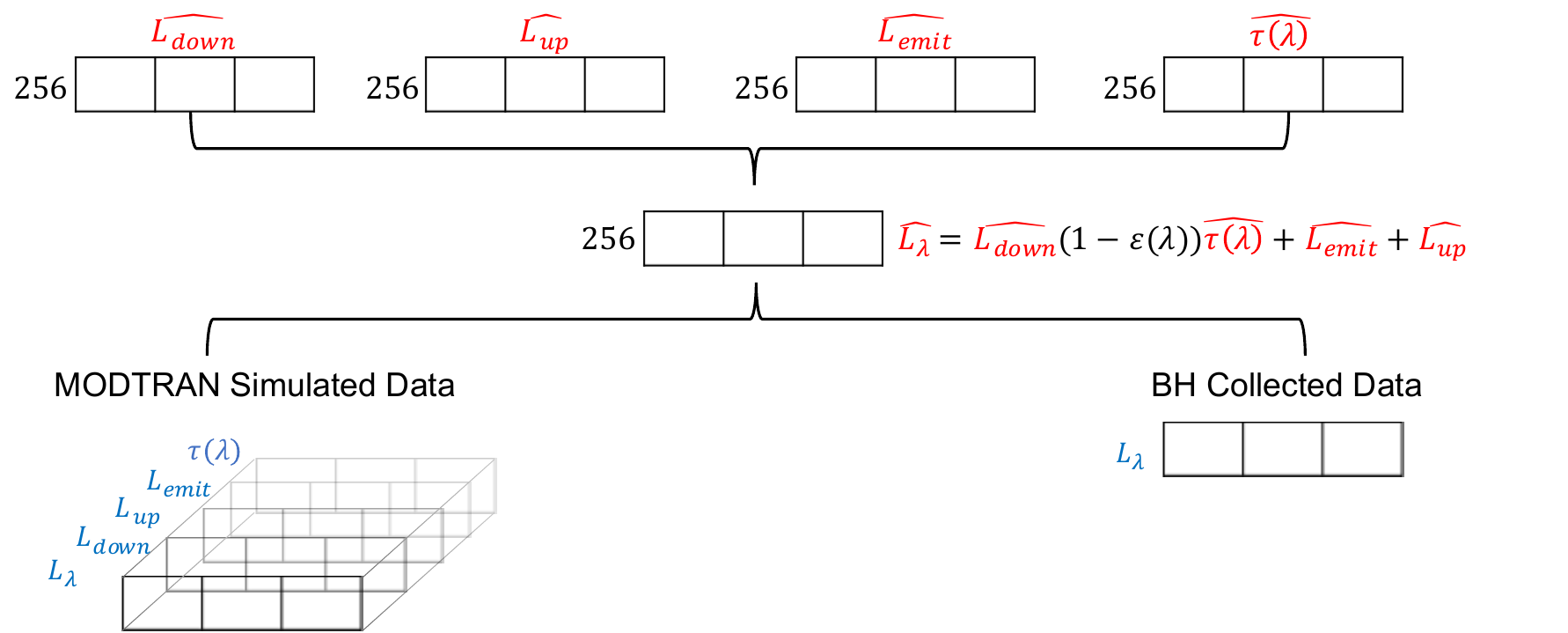}
	\caption{Loss Function For BH and MODTRAN dataset. When the training sample is simulated by MODTRAN, $loss1$ defined in \Eq{equ:lossfunc} is used since ground-truth values of $L_{down}$, $L_{up}$, $L_{emit}$, and $\tau(\lambda)$ are provided. Otherwise, $loss2$ is used for calculated and observed at-sensor total radiance $L_\lambda$.}
	\label{fig:lossfunc}
\end{figure}

\subsubsection{Normalization in the Hybrid Network}
Each ground-truth radiative component (i.e., $L_\lambda$, $L_{down}$, $L_{up}$, $L_{emit}$) is normalized with its global minimum and maximum values which are computed across all materials and all wavelengths during the training process to make sure they are at the same scale of $[0,\ 1]$. The transmission $\tau(\lambda)$ and $\varepsilon(\lambda)$ are within a range of $[0,\ 1]$, which do not need to be normalized as the remaining geometric variables. \Table{tb:(un)norm} illustrates the ground-truth and predicted normalized and unnormalized variables. The minimum and maximum values for normalization are reported in \Table{tb:normalize}.
\begin{table*}[ht]
    \begin{center}
    \caption{Normalized versus unnormalized Components}\label{tb:(un)norm}
    \begin{tabular}{ l|l|l}
    \hline
    & Ground-truth & Predicted/Derived \\\hline
    \bf Normalized & $L_\lambda$, $L_{down}$, $L_{up}$, $L_{emit}$ & $\widehat{L_{down}}$, $\widehat{L_{up}}$ \\\hline
    \bf unnormalized & range, angle, wavelength, $\tau(\lambda)$, $\varepsilon(\lambda)$, $T$ & $\widehat{\tau(\lambda)}$, $\widehat{L_\lambda}$, $\widehat{L_{emit}}$\\\hline
    \end{tabular}
    \end{center}
\end{table*}

\begin{table}[ht]
    \begin{center}
    \caption{Global Minimum and Maximum Values for Normalization}\label{tb:normalize}
    \begin{tabular}{ l|l|l}
    \hline
    & Minimum & Maximum \\\hline
    $L_\lambda$ & $3.656*10^{-04}$ & $1.344*10^{-03}$ \\
    $L_{down}$ & $2.1833*10^{-04}$ & $1.01*10^{-03}$\\
    $L_{up}$ & $1.8016*10^{-04}$ & $7.106*10^{-04}$\\
    $L_{emit}$ & $3.544*10^{-04}$ & $1.116*10^{-03}$\\\hline
    \end{tabular}
    \end{center}
\end{table}

\Eq{equ:simpRTE} only holds when each component is  unnormalized. The network predicted $\widehat{L_{down}}$ and $\widehat{L_{up}}$ are both normalized but the $\widehat{\tau(\lambda)}$ is within $[0,\ 1]$ so it is unnormalized. $\widehat{L_{emit}}$ in \Eq{equ:lossfunc} is calculated by the unnormalized $T$, $\varepsilon(\lambda)$ and predicted $\widehat{\tau(\lambda)}$ so it is also unnormalized. A re-scale process is required for the predicted normalized $\widehat{L_{down}}$ and $\widehat{L_{up}}$ to estimate the unnormalized at-sensor total radiance $\widehat{L_\lambda}$ with unnormalized $\widehat{L_{emit}}$, $\varepsilon(\lambda)$ and $\widehat{\tau(\lambda)}$. When the calculation of $\widehat{L_\lambda}$ and $\widehat{L_{emit}}$ is done, both of them should be re-scaled using their own global minimum and maximum values in \Table{tb:normalize} to further calculate the loss function and do the backpropagation. The expansion of the RTE for normalized components will not be given in this article.

\subsection{Temperature Emissivity Separation}
Given unknown targets, the network should also be able to accurately characterize the atmospheric radiative components that are independent of target properties under different geometries because all the inputs and outputs of the proposed network are target-independent. Only the loss function defined to train the model has introduced the target spectra into calculated $\widehat{L_{emit}}$ and and $\widehat{L_\lambda}$. This means the prediction of downwelling $\widehat{L_{down}}$, upwelling $\widehat{L_{up}}$ and transmission $\widehat{\tau(\lambda)}$ should not be affected whether the targets are known or unknown after the network is well-trained with known materials.

Once all target-independent radiative components (i.e., $\widehat{L_{down}}$, $\widehat{L_{up}}$, $\widehat{\tau(\lambda)}$) can be predicted accurately via the network, the target-involved contributions to at-sensor total radiance can be quantified in the next step. In other words, the goal becomes to use the observed at-sensor total radiance and predicted atmospheric components to estimate the target temperature and spectral emissivity. \Eq{equ:simpRTE} and \Eq{equ:Planck} suggest that there are $n+1$ unknown variables ($n$ emissivity + $1$ target temperature), given the radiance measured at $n$ spectral wavelength bands. Extra information is needed to solve the target spectral emissivity and temperature. In remote sensing, this is called the Temperature Emissivity Separation (TES) problem.

A grid search method is proposed to estimate the emissivity by giving a range of evenly distributed temperatures (e.g., $[280, 320]$ by every $5\ (Kelvins)$) and an assumed constant emissivity value $\epsilon$ (typically the mean value of 256 spectral emissivity) for each material. For example, the $\epsilon$ of water or green healthy grass are usually be around $0.98-0.99$. For a specific temperature $T$, the spectral emissivity can be calculated as:
\begin{equation}\label{equ:emissEq}
    \begin{split}
     & (1-\varepsilon(\lambda))L_{down} + \varepsilon(\lambda)L_{T\lambda} = \frac{L_\lambda - L_{up}}{\tau(\lambda)}\\
     & \varepsilon(\lambda) = \frac{(L_\lambda - L_{up})/\tau(\lambda) - L_{down}}{L_{T\lambda}-L_{down}}
    \end{split}
\end{equation} 
It is possible to estimate the target temperature where its corresponding retrieved emissivity has a smallest MAE with the assumed $\epsilon$ value.

\section{Results and Analysis}
To demonstrate the generalization ability of the proposed network, the evaluation is performed on both MODTRAN and BH test datasets. The mean absolute error (MAE) and root mean squared error (RMSE) are computed over spectra, angles, ranges and materials for: 1) each predicted atmospheric radiative component (i.e., $\widehat{L_{down}}$, $\widehat{L_{up}}$, $\widehat{\tau(\lambda)}$) for MODTRAN test dataset, 2) at-sensor target emittance ($\widehat{L_{emit}}$) derived based on \Eq{equ:lossfunc} for MODTRAN test dataset, and 3) at-sensor total radiance ($\widehat{L_\lambda}$) derived based on \Eq{equ:lossfunc} for both MODTRAN and BH test dataset.

For the surfaces of known emissivity $\varepsilon(\lambda)$ and temperature $T$, all predicted or derived radiative components have errors at least one-order of magnitude smaller than the ground-truth signatures in 1)-3) evaluations above. For unknown surfaces, the network is also able to predict all atmospheric components with the same errors as targets with a known $\varepsilon(\lambda)$ and temperature $T$ in 1) evaluation. The RMSE of retrieved surface emissivity at the estimated temperature using the grid search method, is calculated as another indicator of the network ability in both MODTRAN and BH test dataset.

\subsection{Atmospheric Correction}
To validate the accuracy of the atmospheric correction of the proposed hybrid network, the temperature $T$ and spectral emissivity $\varepsilon(\lambda)$ of all target surfaces in the testing dataset are masked as 0. \Table{tb:errorComps} shows the ground-truth value of each radiative component as well as the MAE and RMSE of its predictions averaged for all 29 materials under 1116 different geometries over 256 wavelengths with the true and false target temperature and emissivity provided.

\begin{table*}[ht]
    \centering
	\caption{MAE and RMSE with true and false target temperature and spectral emissivity}
	\label{tb:errorComps}
    \begin{tabular}{l|l|l|l|l|l}
    \hline
    \bf{Components} & $\widehat{L_\lambda}$ & $\widehat{L_{down}}$ & $\widehat{L_{up}}$ & $\widehat{L_{emit}}$ & $\widehat{\tau(\lambda)}$\\
    \hline
    \diagbox[height=2.5em, width=9em]{\bf{Errors}}{\bf{Mean}} &
    $8.44*10^{-4}$ & $4.07*10^{-4}$ & $3.72*10^{-4}$ & $4.44*10^{-4}$ & $4.86*10^{-1}$\\
    \hline
    MAE (True) & $2.84*10^{-6}$ & $4.65*10^{-6}$ & $1.72*10^{-6}$ & $3.77*10^{-6}$ & $3.61*10^{-3}$\\
    MAE (False) & $4.74*10^{-4}$ & $4.65*10^{-6}$ &  $1.72*10^{-6}$ & $4.44*10^{-4}$ & $3.61*10^{-3}$ \\
    \hline
    RMSE (True) & $3.45*10^{-6}$ &  $9.17*10^{-6}$ & $2.06*10^{-6}$ & $4.35*10^{-6}$ & $4.23*10^{-3}$ \\
    RMSE (False) & $5.22*10^{-4}$ & $9.17*10^{-6}$ & $2.06*10^{-6}$ & $4.92*10^{-4}$ & $4.23*10^{-3}$ \\
    \hline
    \end{tabular}
\end{table*}

It is possible to see, the ground-truth radiative components vary for different materials and geometries. The mean value is around $10^{-4}$ order of magnitude except for the atmospheric transmission whose range is between 0 and 1. When the correct target temperature and spectral emissivity are provided, both MAE and RMSE are at least 2 orders of magnitude smaller than the original signature. When the target is unknown, the target-independent atmospheric components (i.e., $\widehat{L_{down}}$, $\widehat{L_{up}}$, and $\widehat{\tau(\lambda)}$), are predicted with the same accuracy. However, the MAE and RMSE of $\widehat{L_{emit}}$ and $\widehat{L_\lambda}$ are two orders of magnitude larger when the false $T$ and $\varepsilon(\lambda)$ are provided, which is of the same magnitude as the original signature. This means our proposed network can predict the atmospheric components accurately under different geometries no matter whether the target information is available or not. 

\begin{figure}[tp]
\begin{tabular}{cc}
\includegraphics[width=0.47\textwidth]{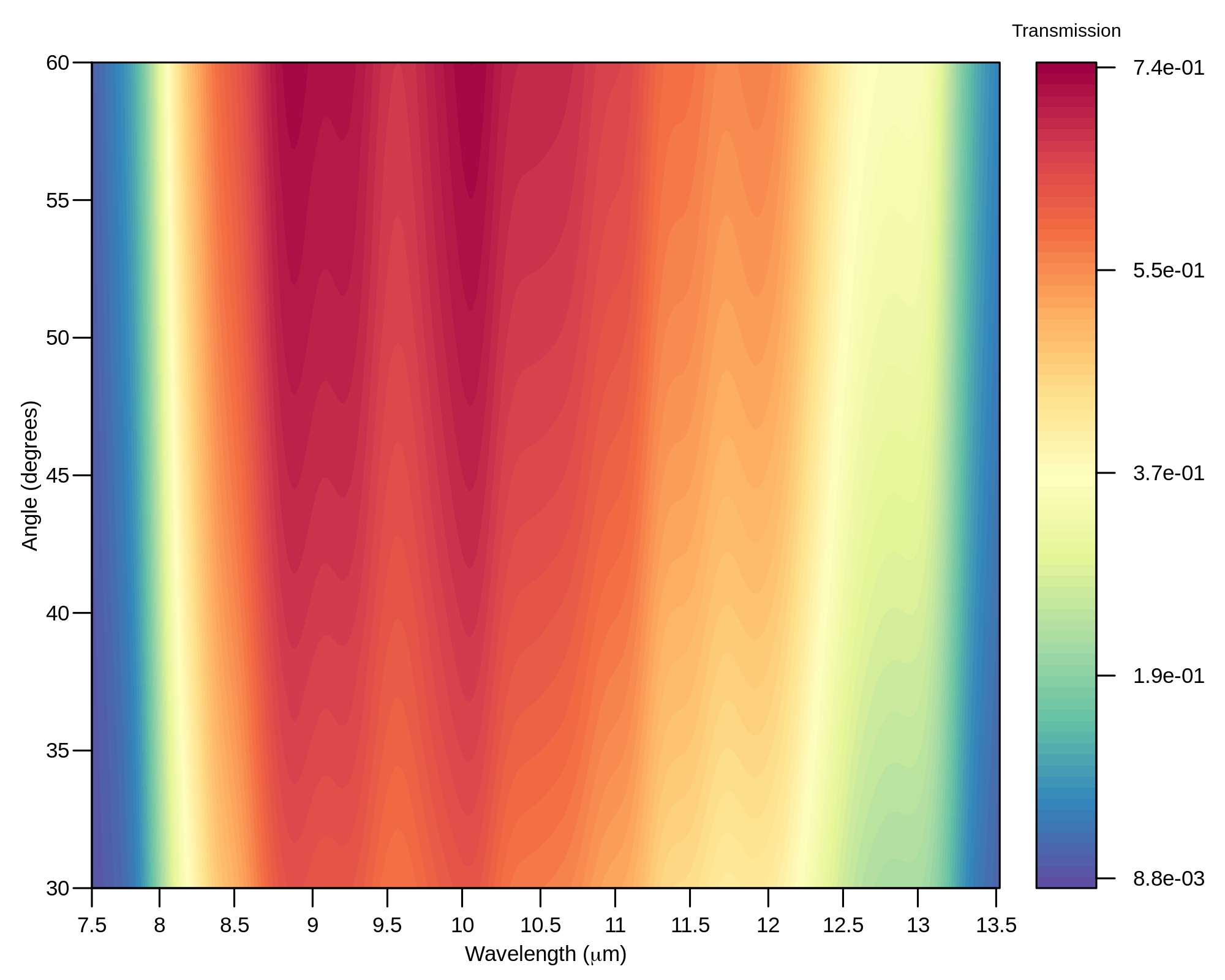} & \includegraphics[width=0.47\textwidth]{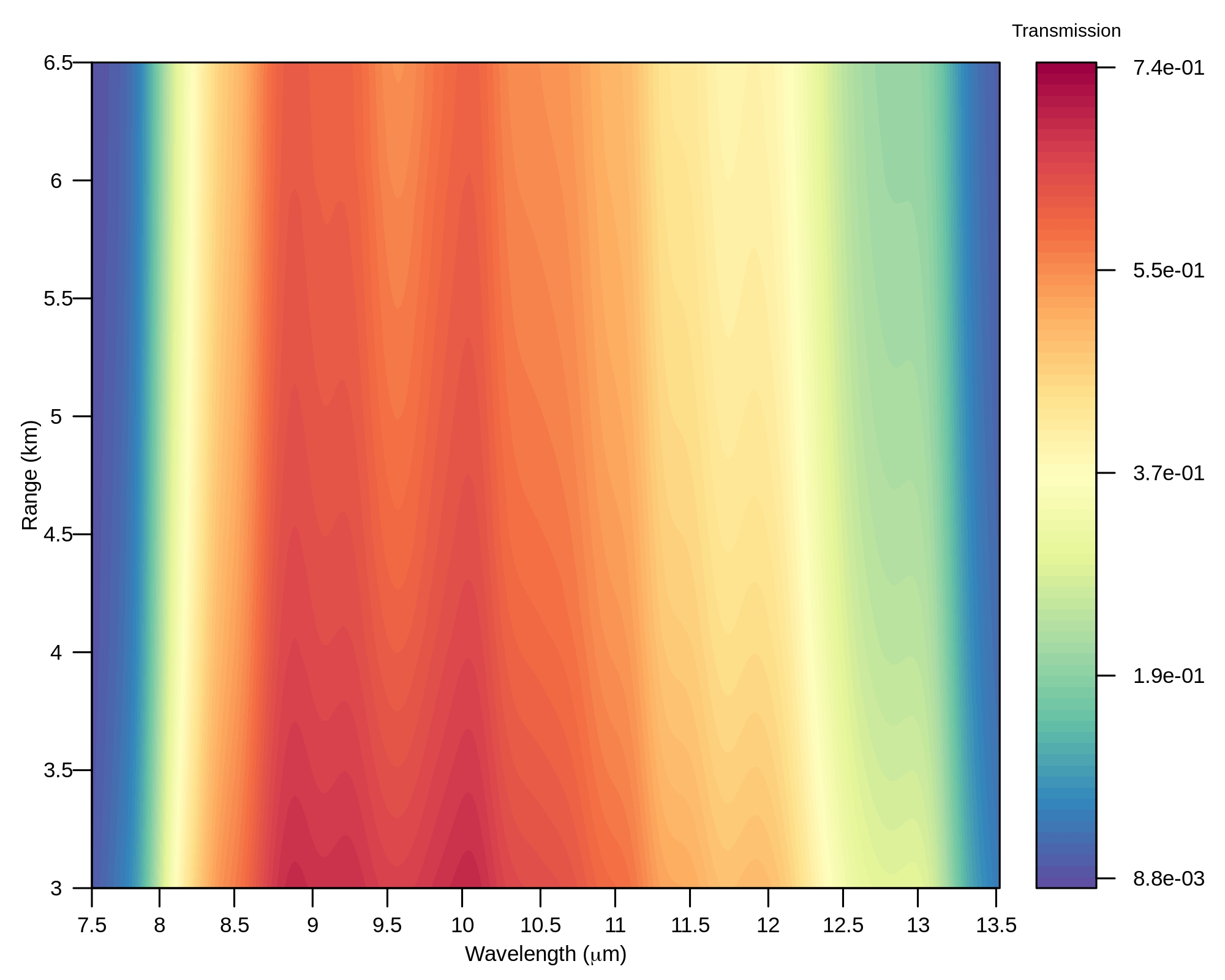}\\
\kern 10pt (a1) & \kern 10pt (b1)\\
\includegraphics[width=0.47\textwidth]{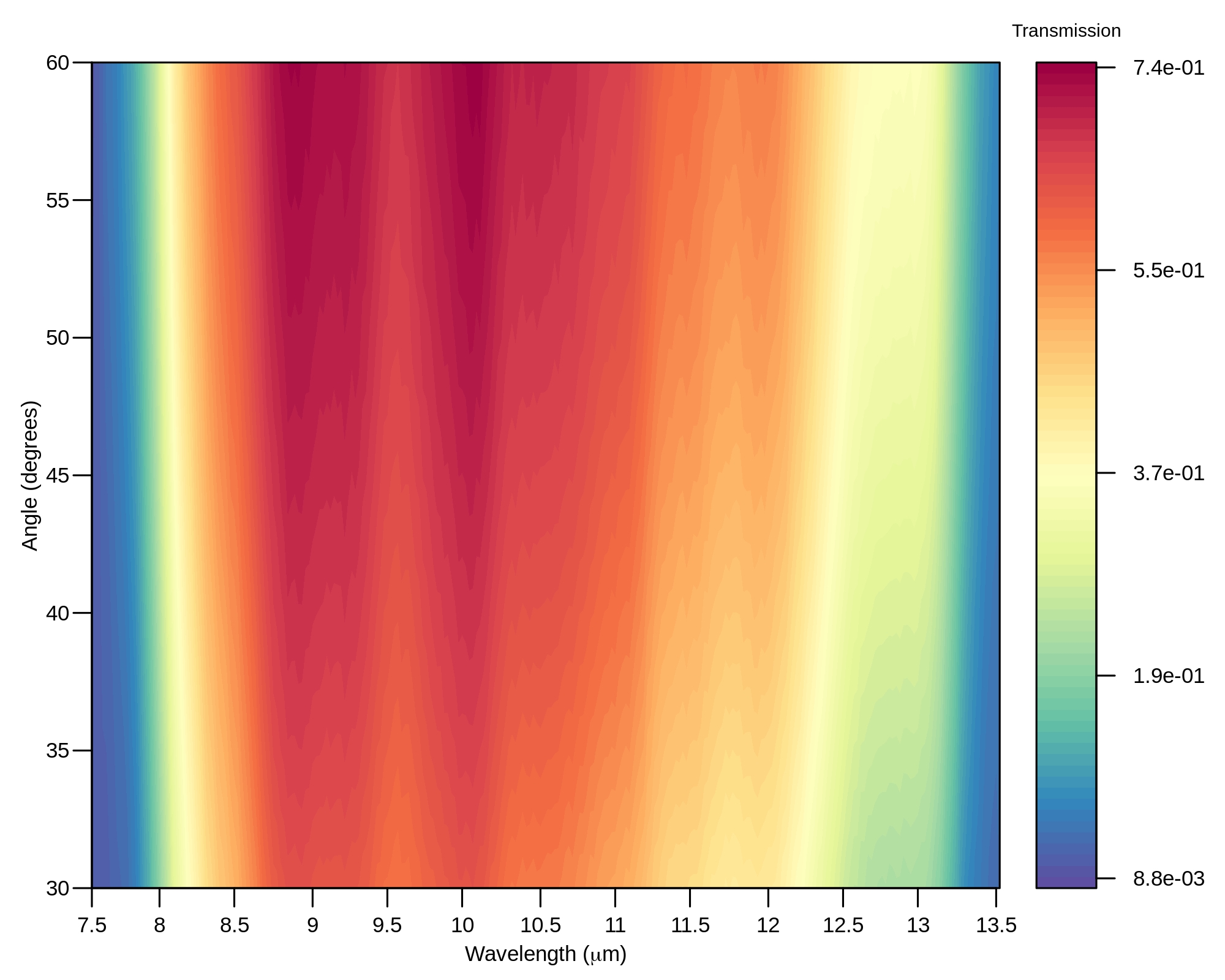} & \includegraphics[width=0.47\textwidth]{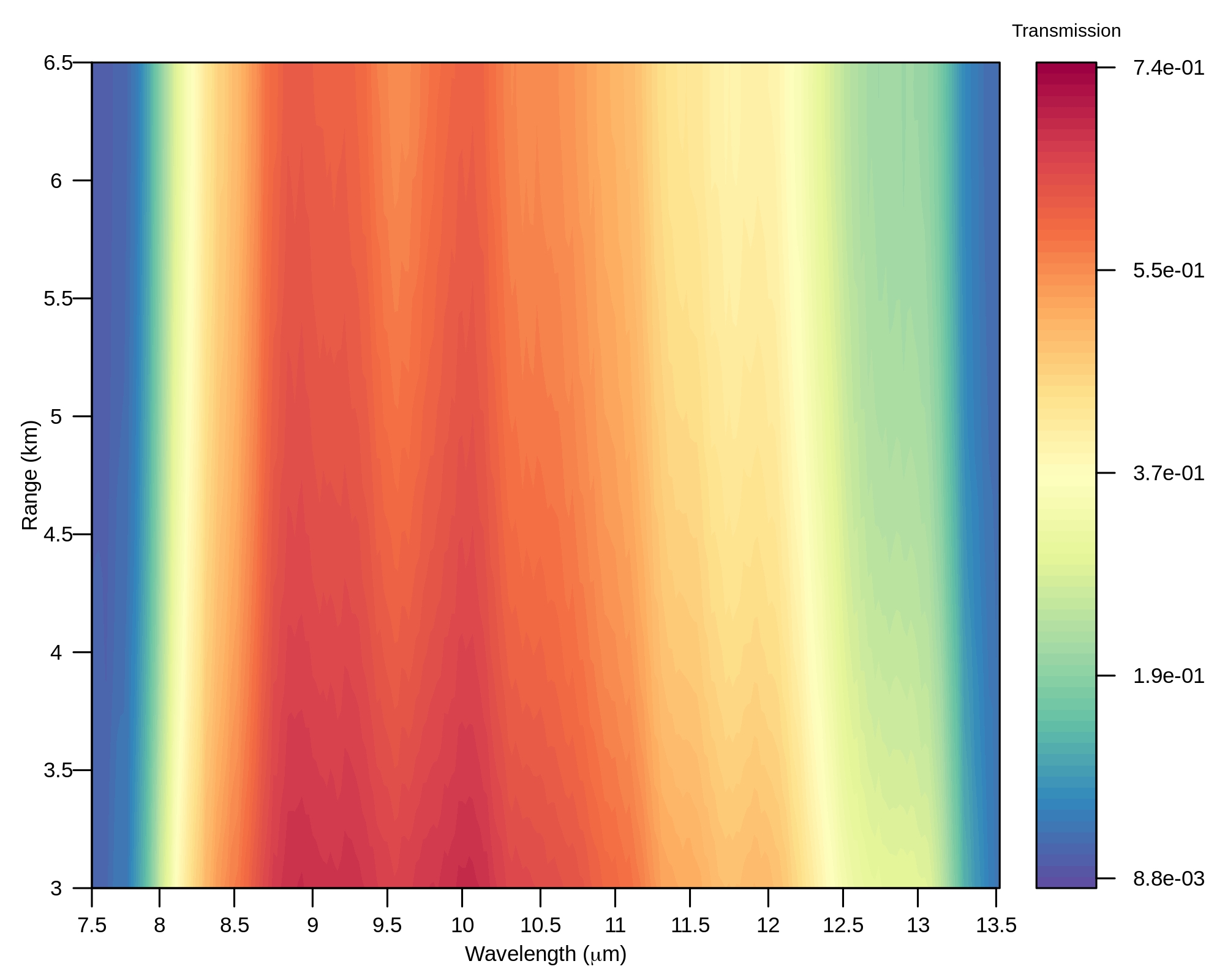}\\
\kern 10pt (a2) & \kern 10pt (b2)\\
\includegraphics[width=0.47\textwidth]{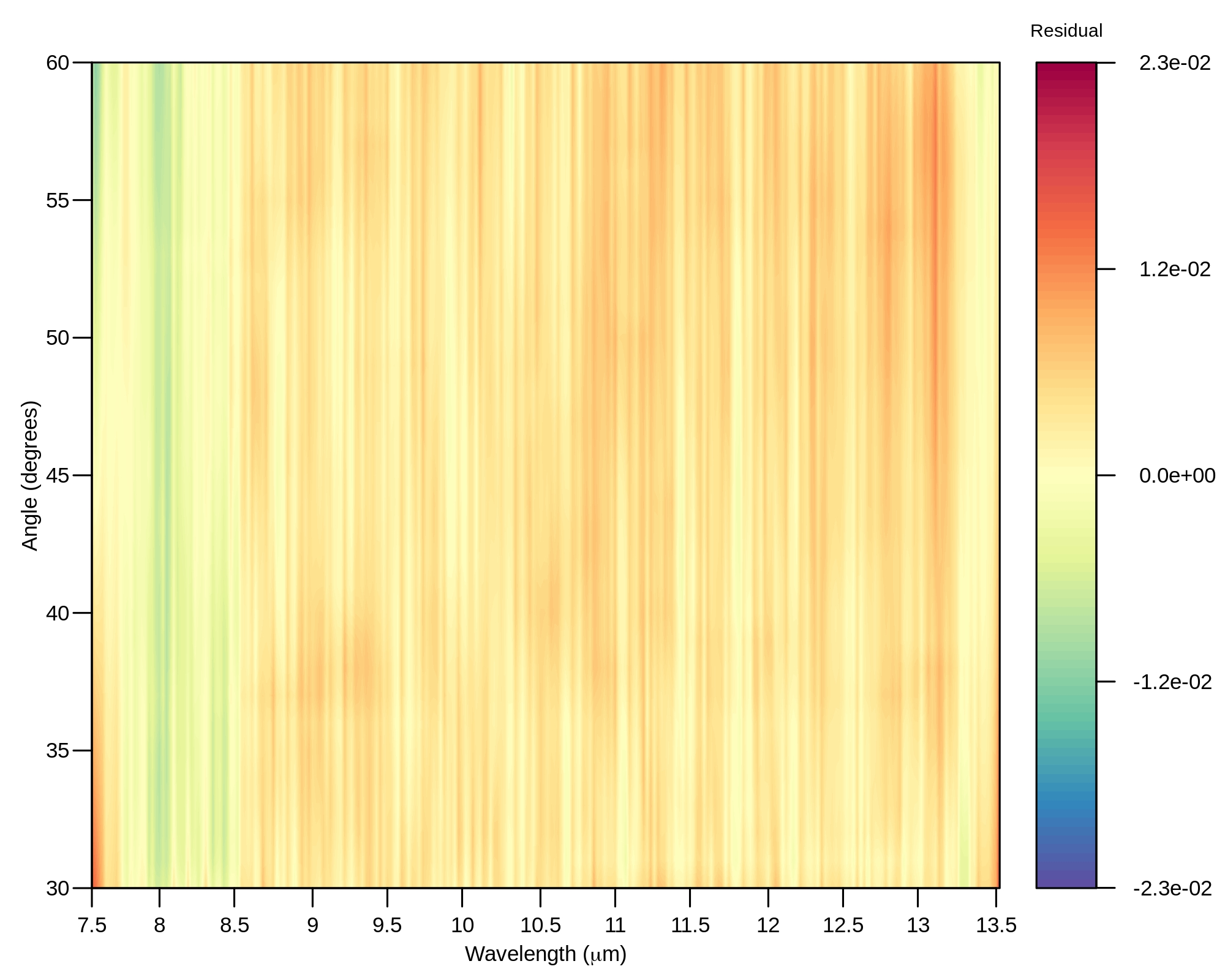} &
\includegraphics[width=0.47\textwidth]{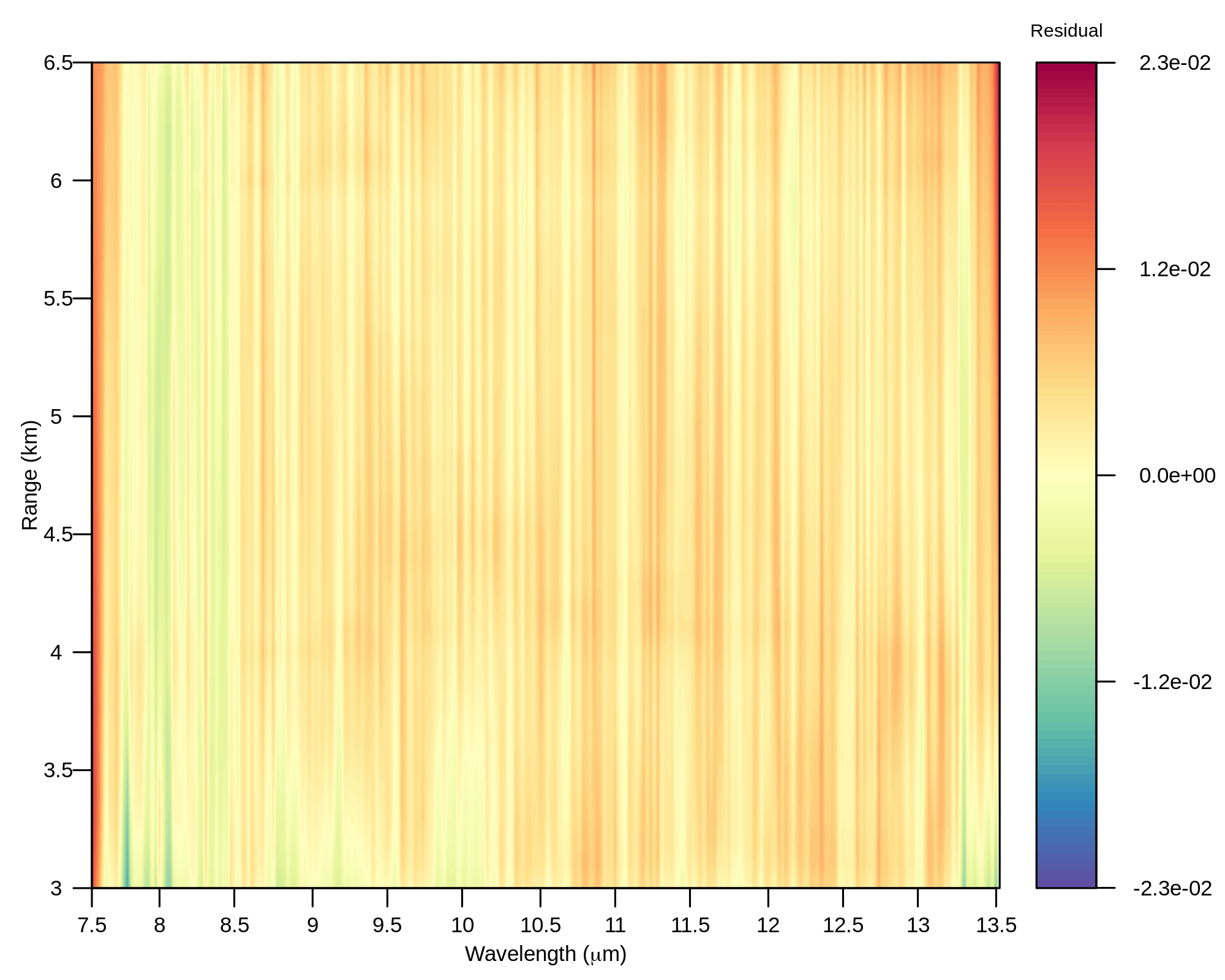}\\
\kern 10pt (a3) & \kern 10pt (b3)\\
\end{tabular}
\caption{Ground-truth and predicted atmospheric transmission under different geometries and their residuals (Aluminum)}
\label{fig:transResidual}
\end{figure}

\begin{figure}[tp]
\begin{tabular}{cc}
\includegraphics[width=0.47\textwidth]{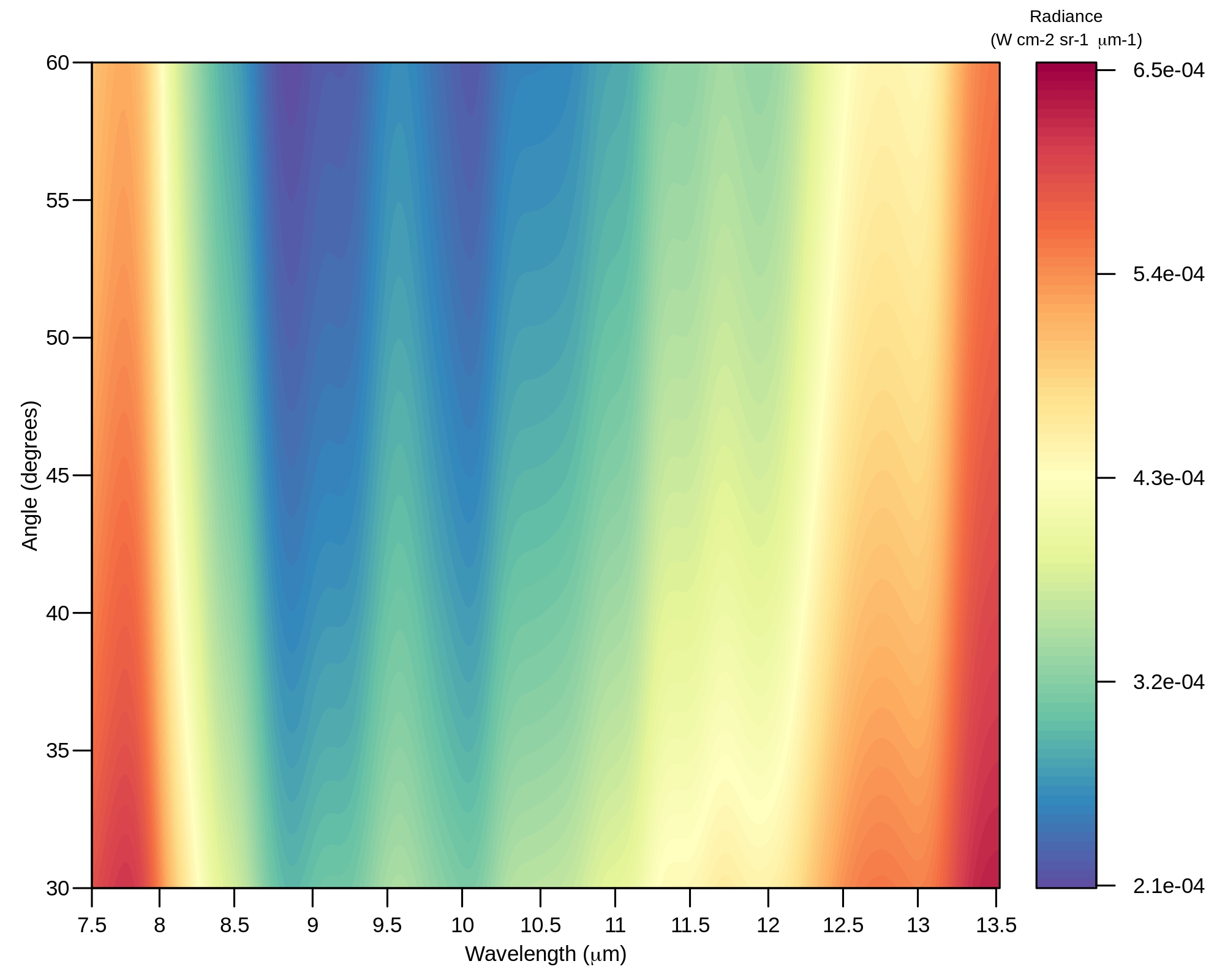} & \includegraphics[width=0.47\textwidth]{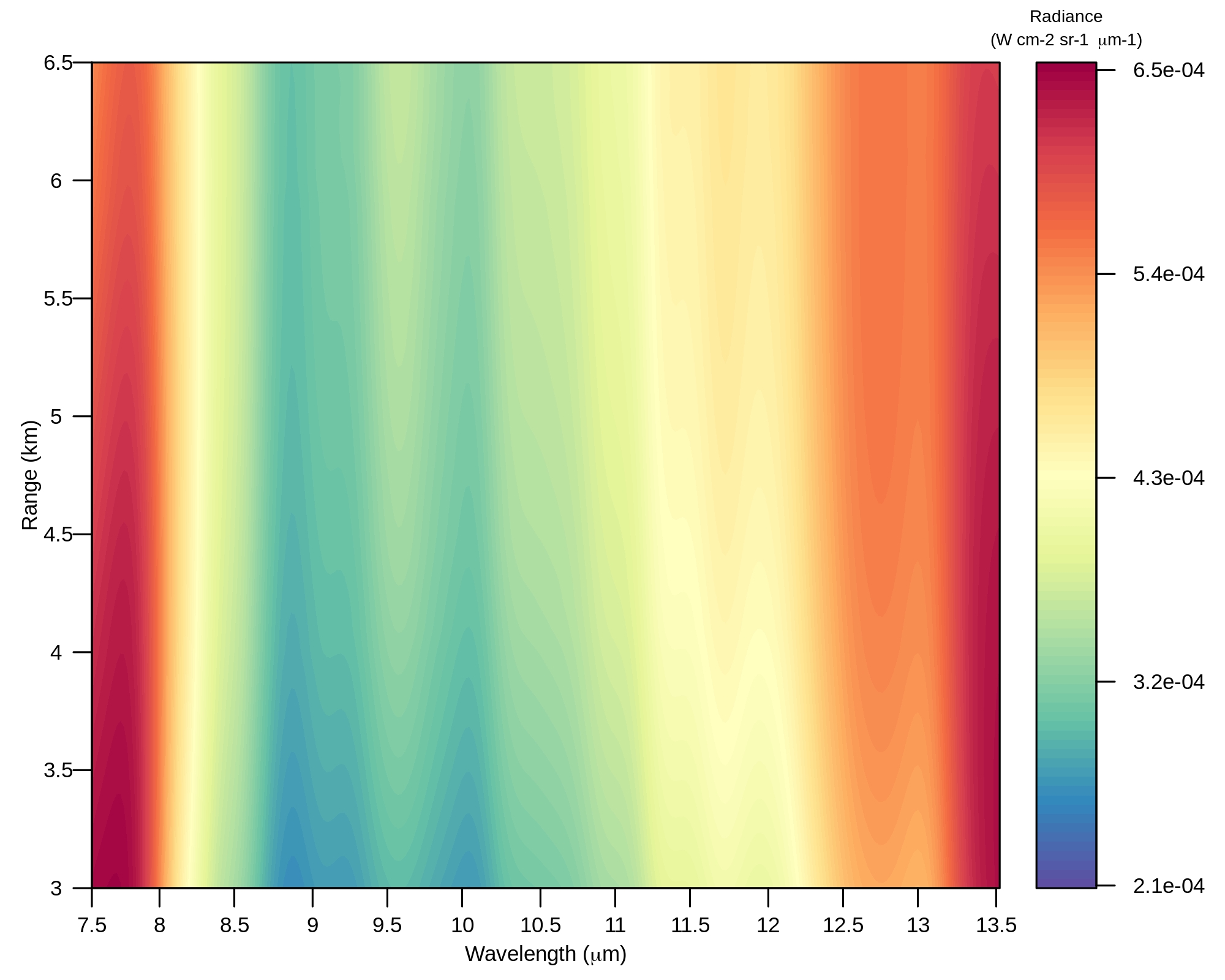}\\
\kern 10pt (a1) & \kern 10pt (b1)\\
\includegraphics[width=0.47\textwidth]{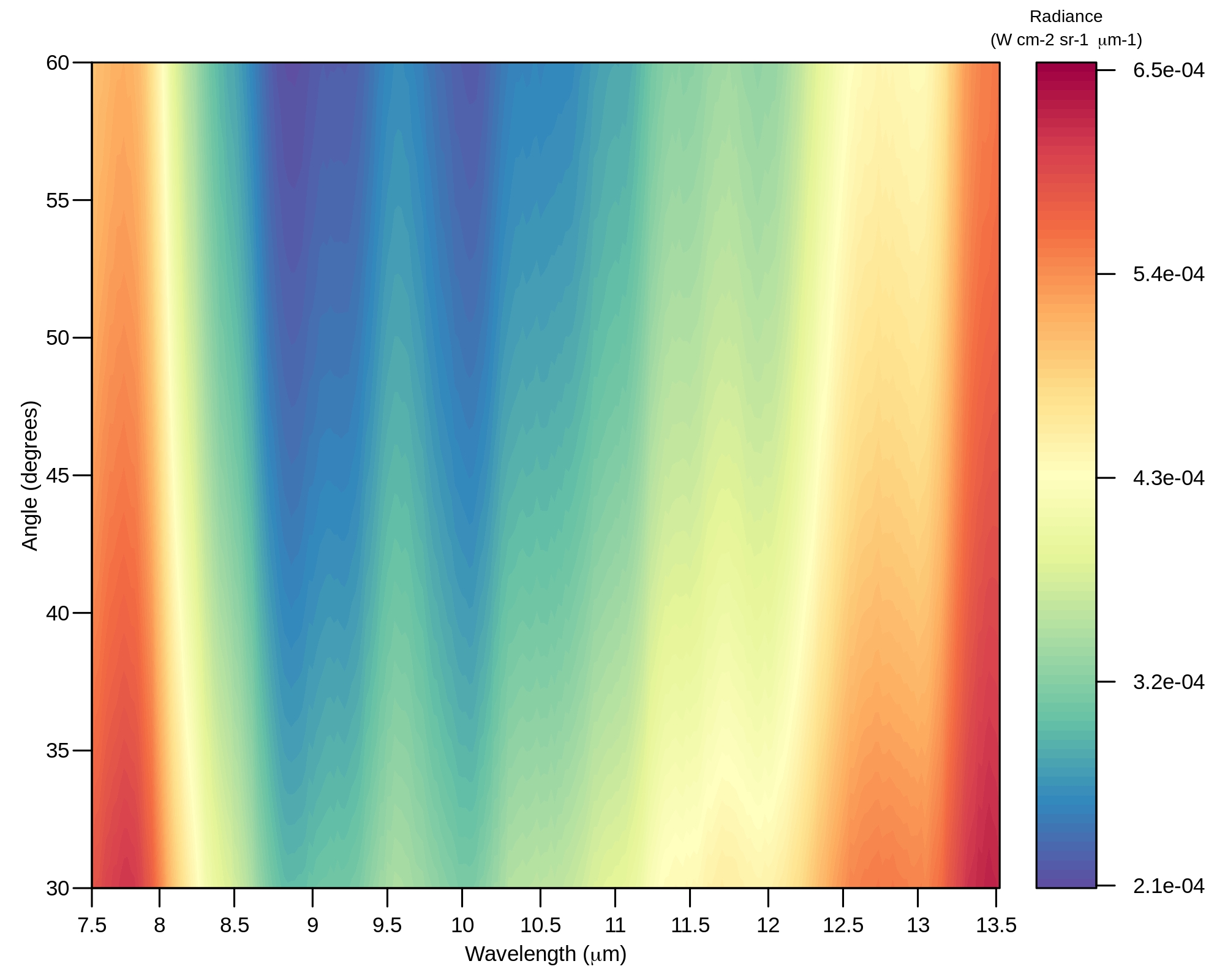} & \includegraphics[width=0.47\textwidth]{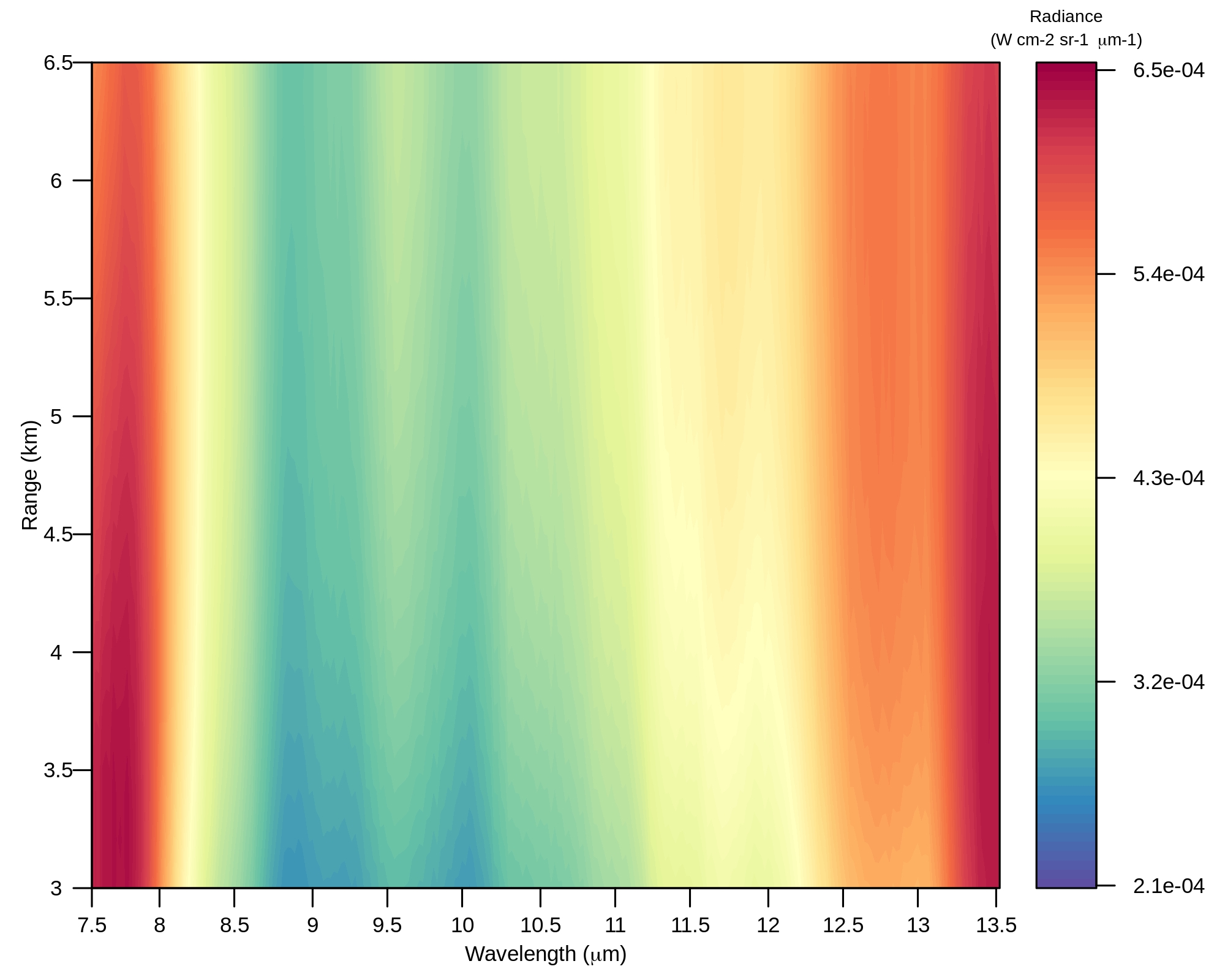}\\
\kern 10pt (a2) & \kern 10pt (b2)\\
\includegraphics[width=0.47\textwidth]{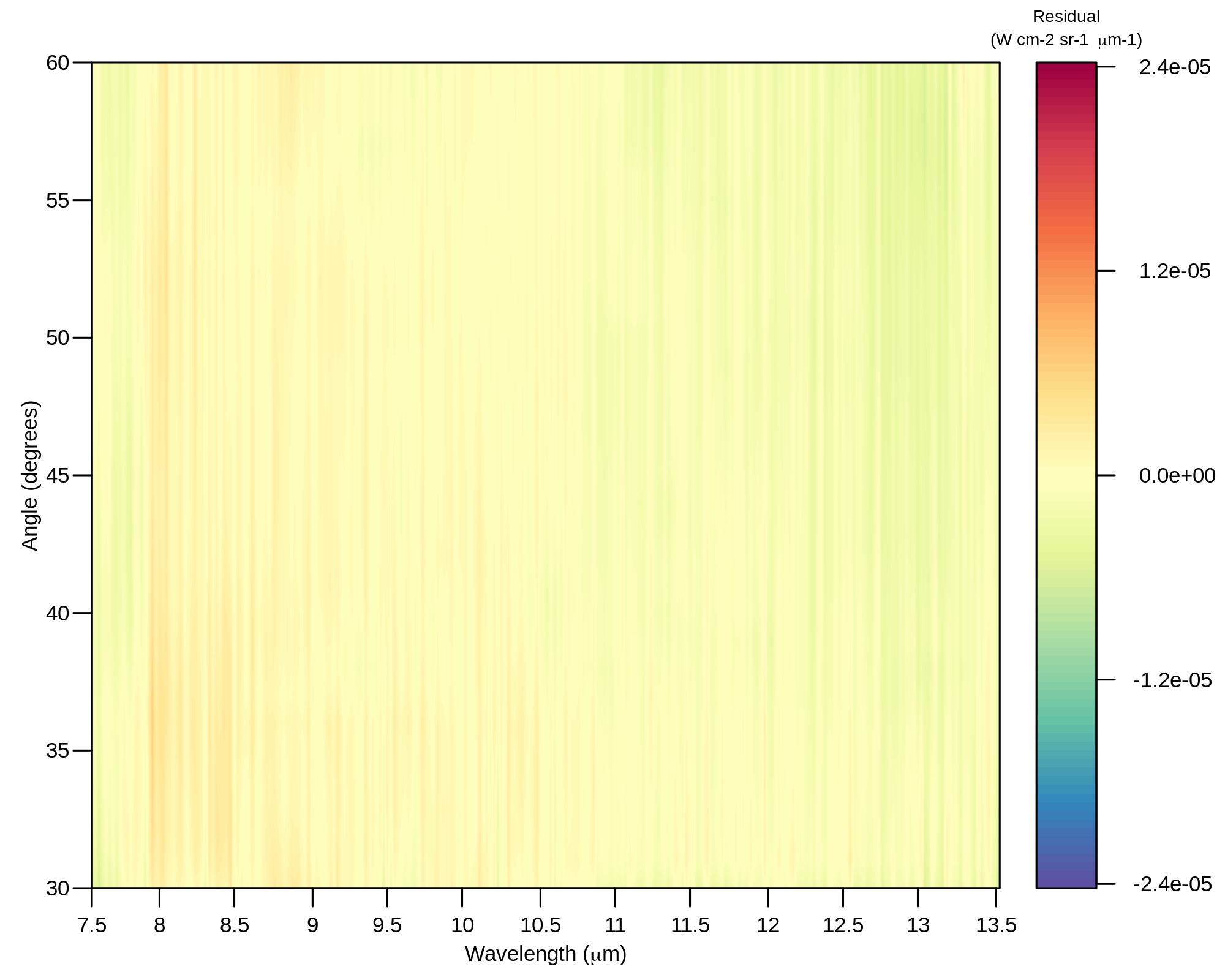} &
\includegraphics[width=0.47\textwidth]{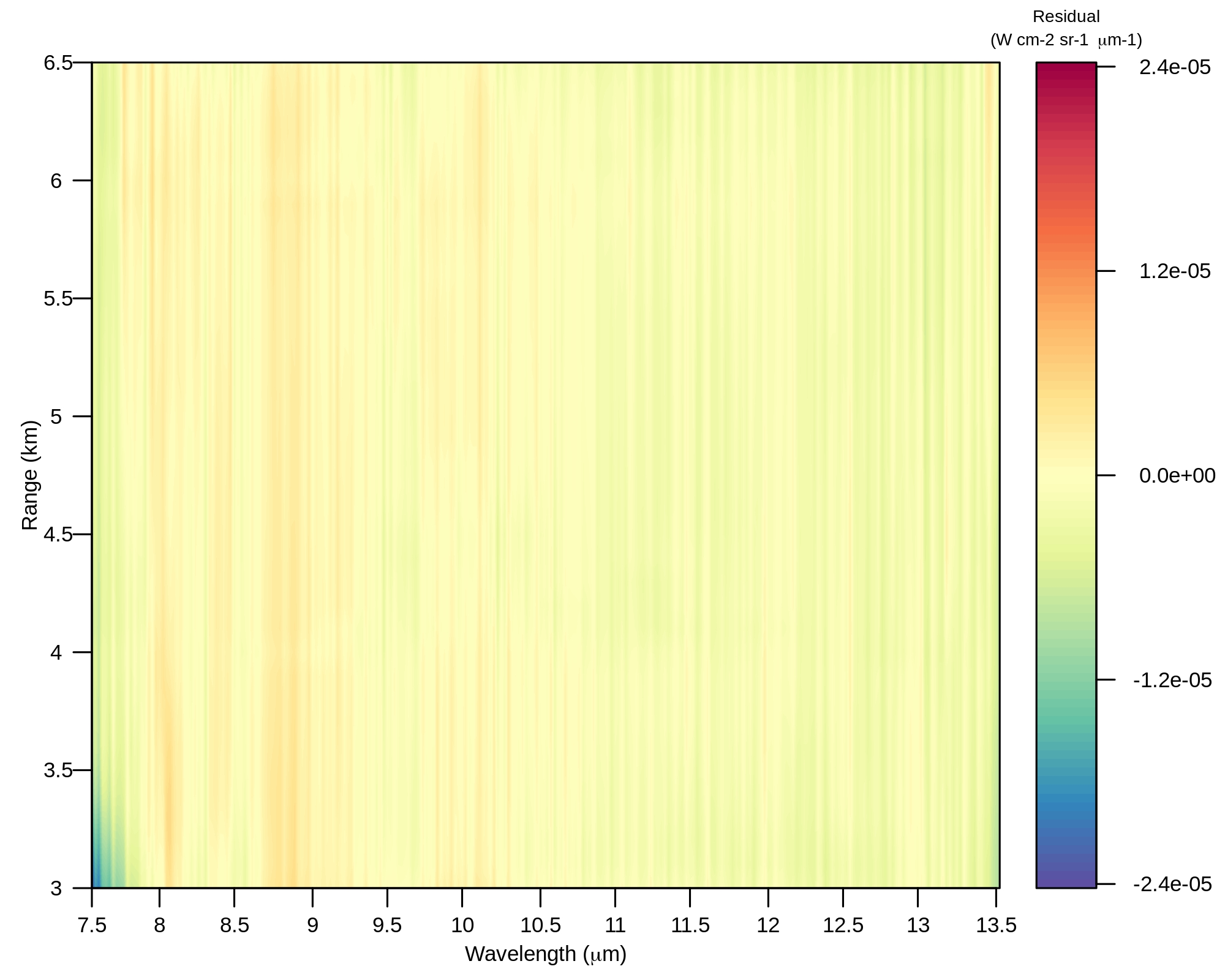}\\
\kern 10pt (a3) & \kern 10pt (b3)\\
\end{tabular}
\caption{Ground-truth and predicted atmospheric upwelling radiance under different geometries and their residuals (Aluminum)}
\label{fig:upResidual}
\end{figure}

\begin{figure}[tp]
\begin{tabular}{cc}
\includegraphics[width=0.47\textwidth]{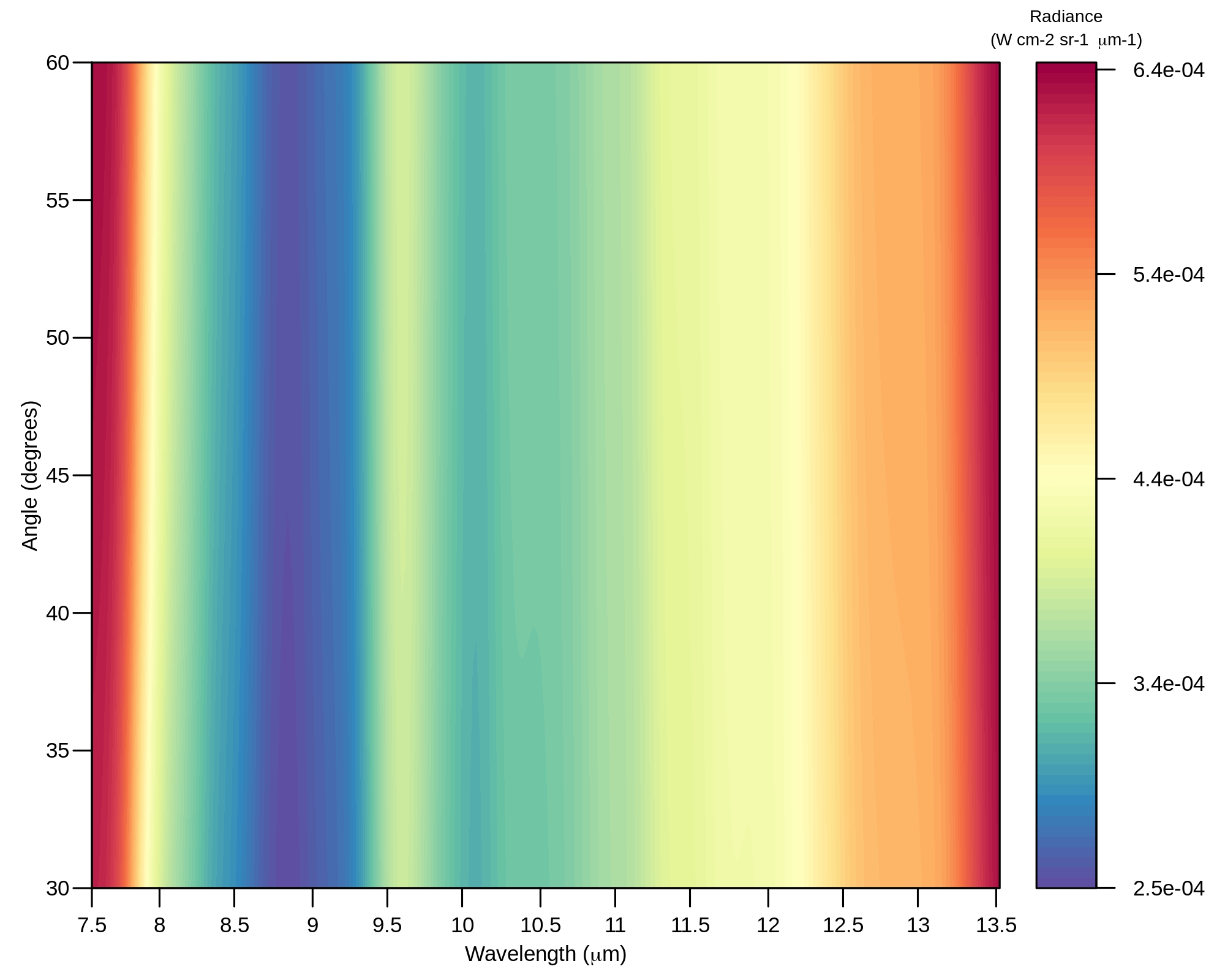} & \includegraphics[width=0.47\textwidth]{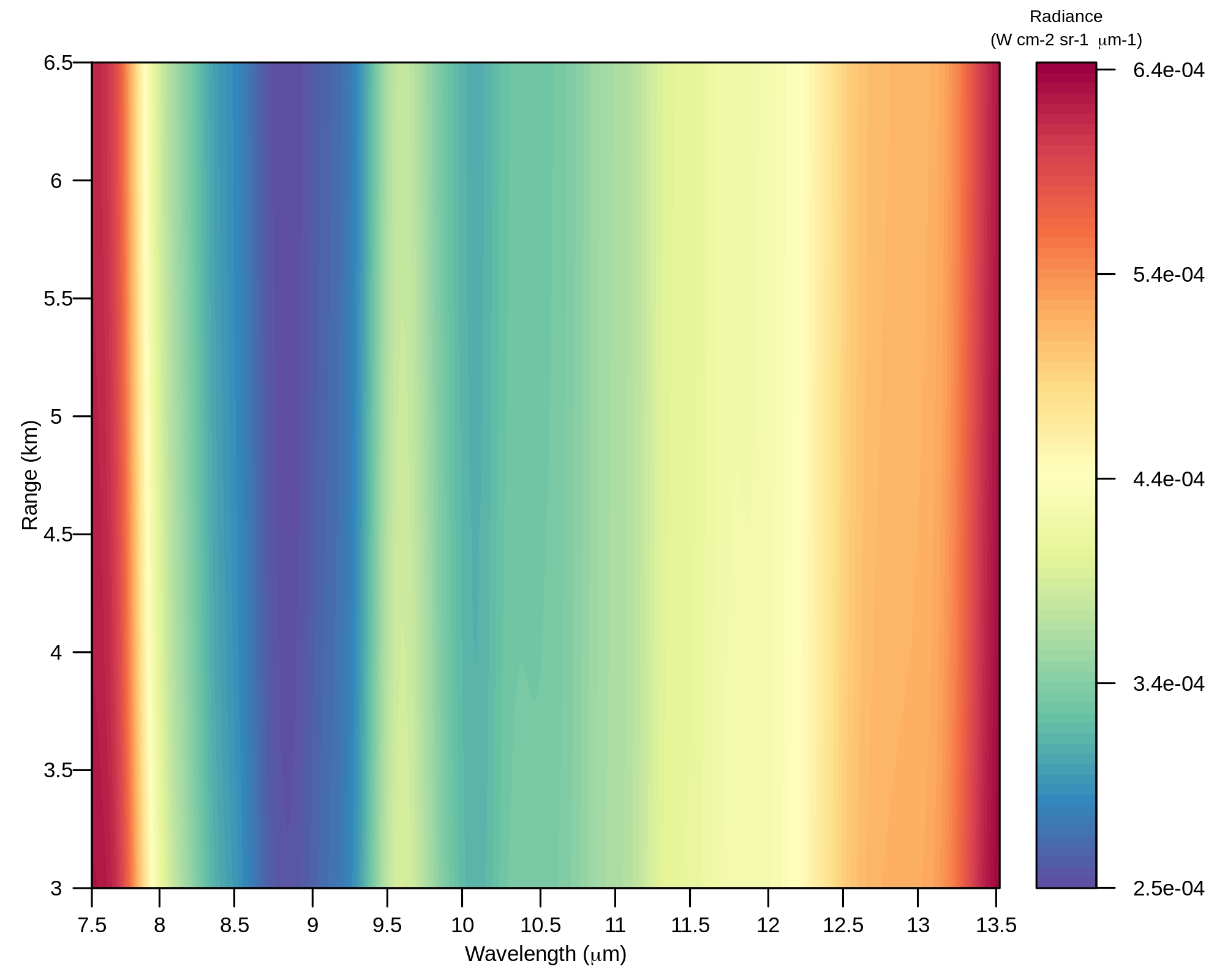}\\
\kern 10pt (a1) & \kern 10pt (b1)\\
\includegraphics[width=0.47\textwidth]{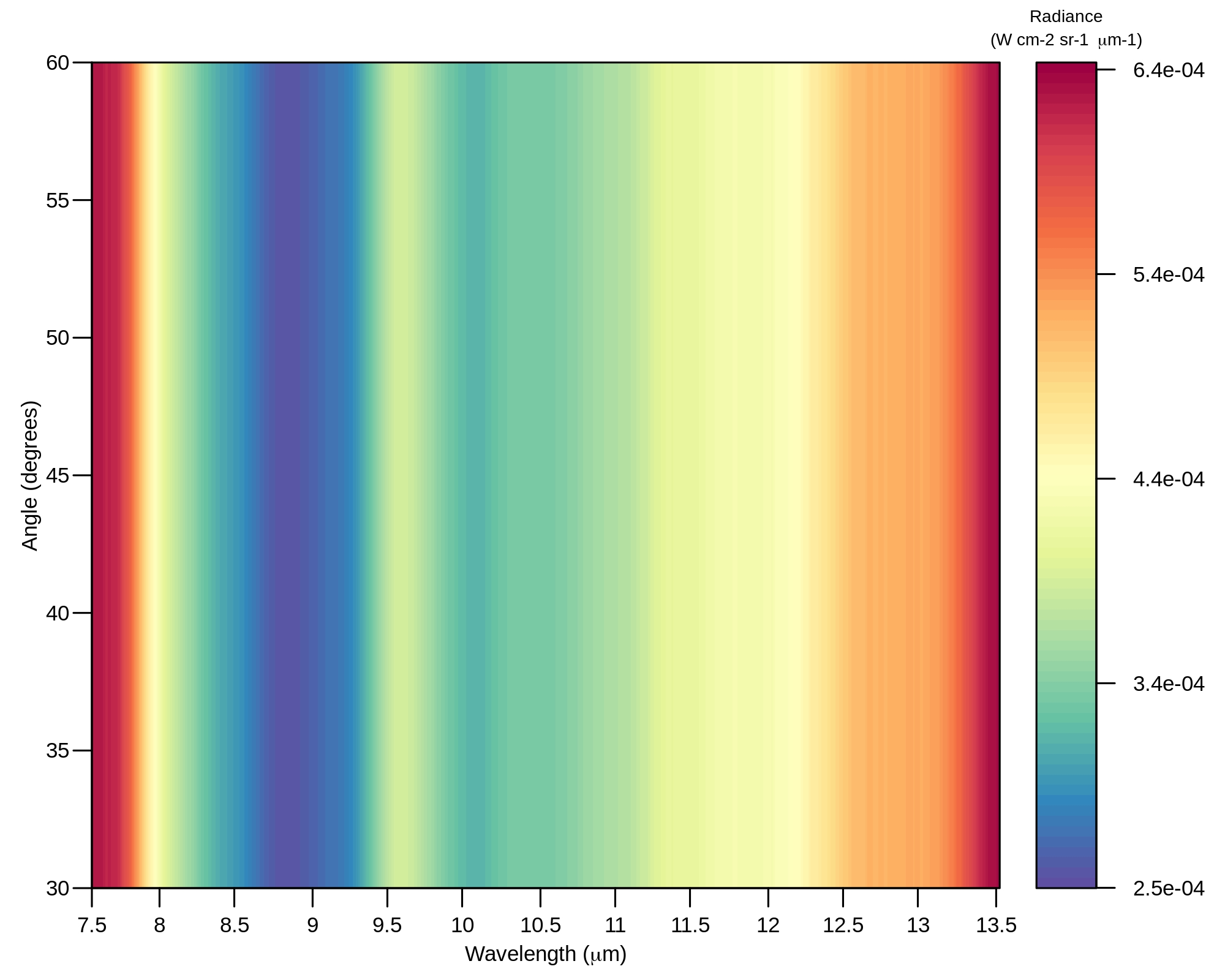} & \includegraphics[width=0.47\textwidth]{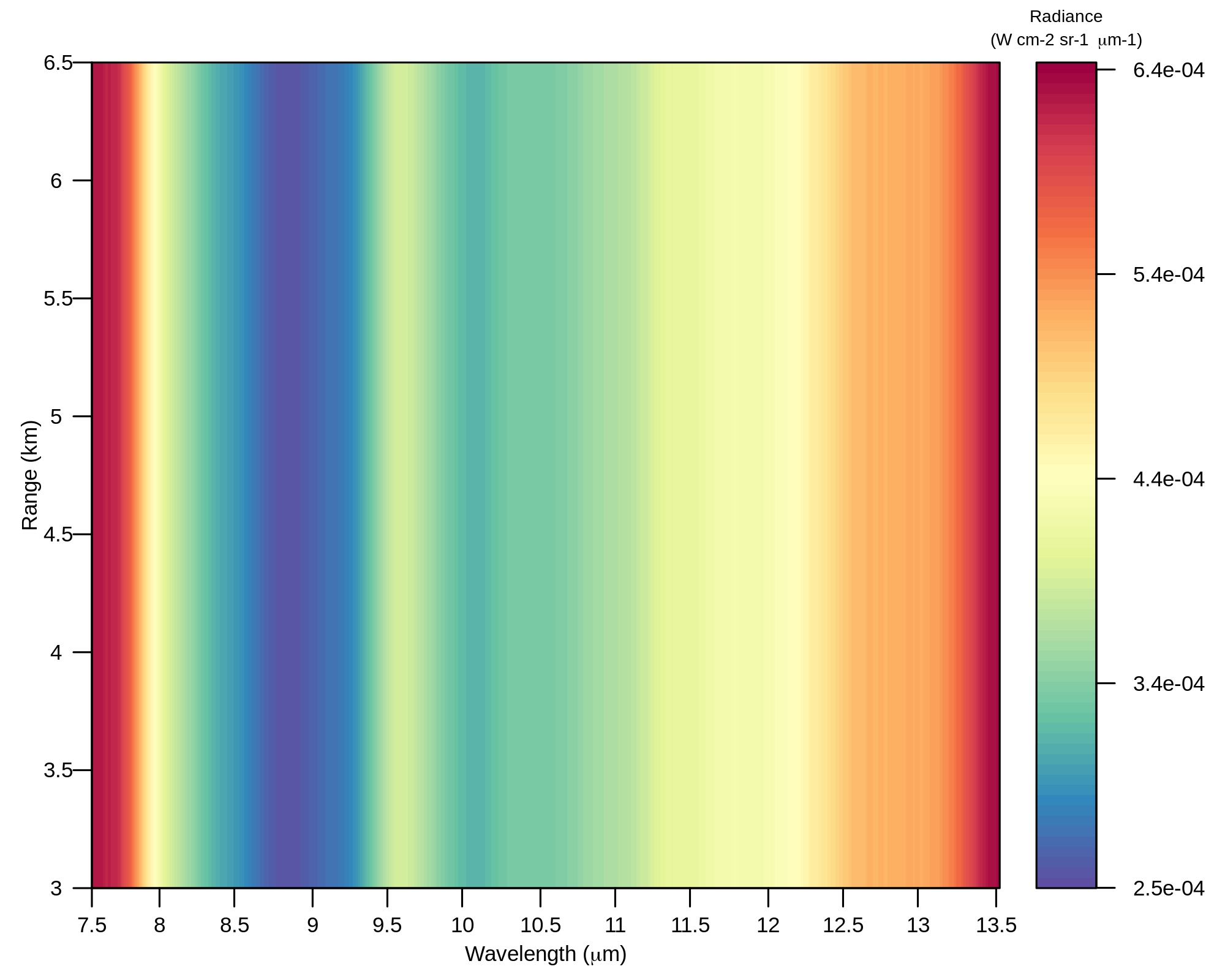}\\
\kern 10pt (a2) & \kern 10pt (b2)\\
\includegraphics[width=0.47\textwidth]{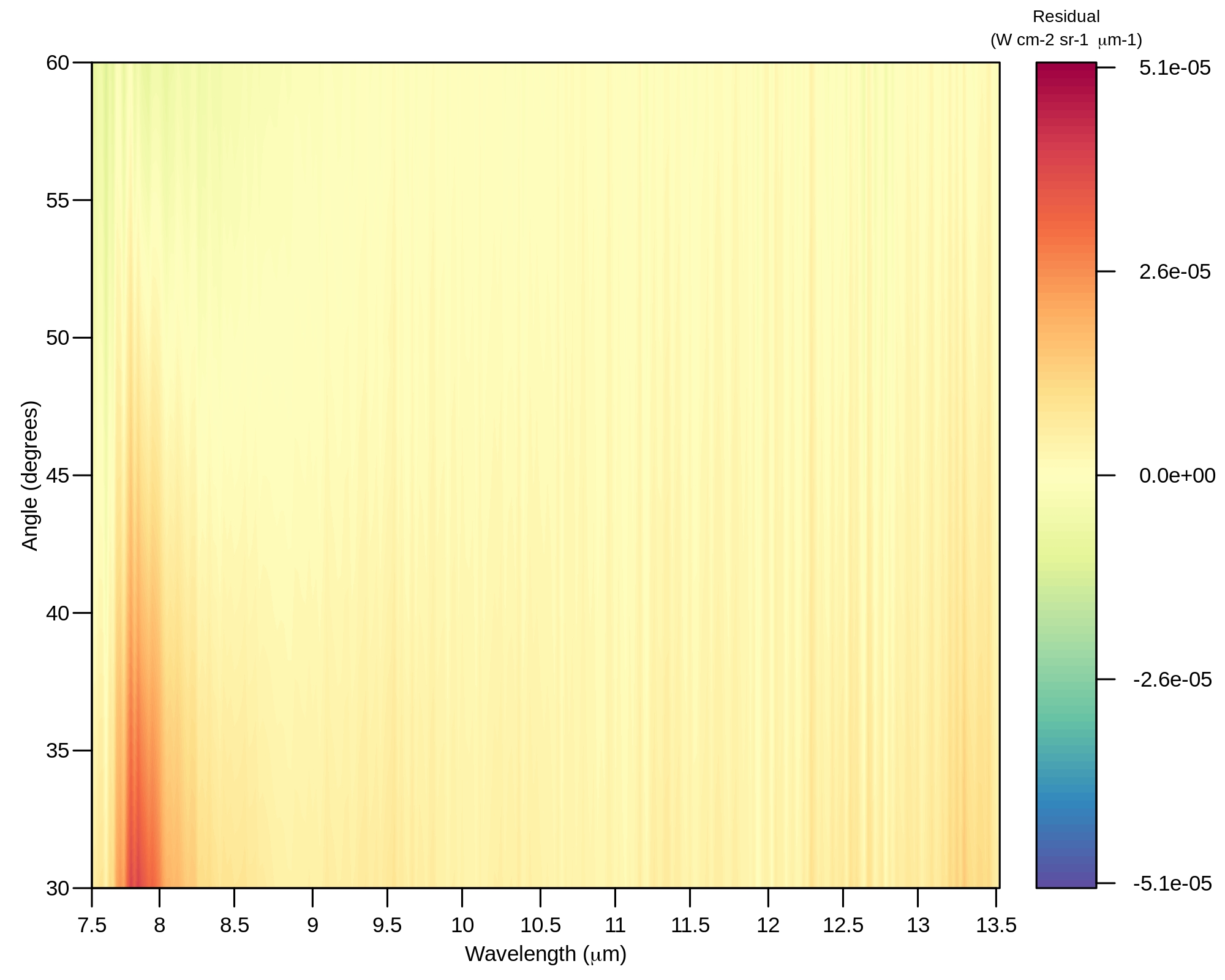} &
\includegraphics[width=0.47\textwidth]{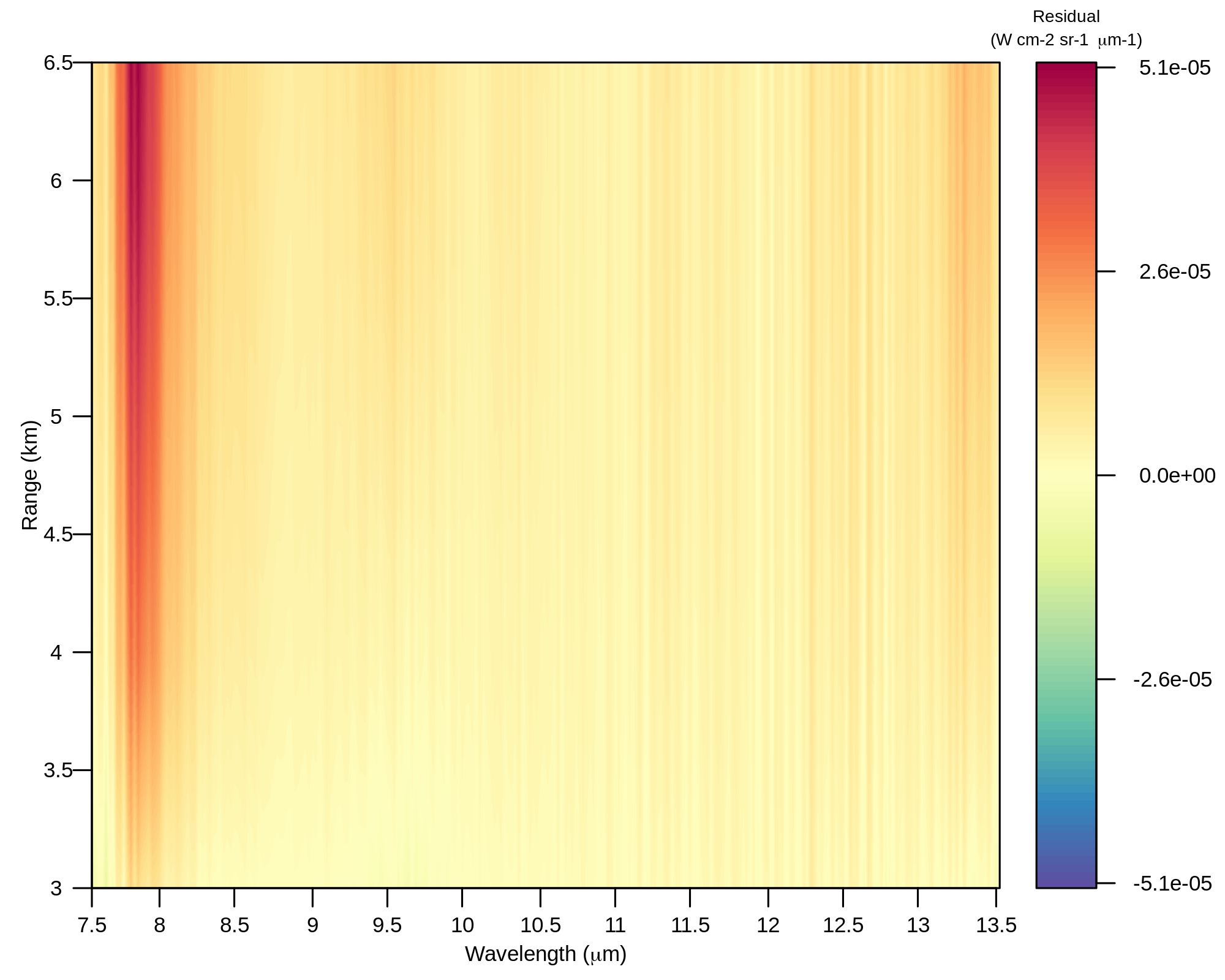}\\
\kern 10pt (a3) & \kern 10pt (b3)\\
\end{tabular}
\caption{Ground-truth and predicted atmospheric downwelling radiance under different geometries and their residuals (Aluminum)}
\label{fig:downResidual}
\end{figure}

\Figs{fig:transResidual}, \ref{fig:upResidual} and  \ref{fig:downResidual} show the ground-truth and predicted atmospheric transmission, upwelling and downwelling radiances as well as their residuals for the Aluminum under different geometries. In these figures, all the plots indicated with ($*1$), ($*2$), and ($*3$) represent the ground-truth, predicted and residual, respectively. Each atmospheric component is shown in (a$*$) as a function of wavelength and elevation angle with a fixed range of $5000\ m$ while in (b$*$) as a function of wavelength and range with a fixed angle of $30^\circ$.

Several wavelength bands have a strong atmospheric absorption on the radiance signatures, shown in \Fig{fig:transResidual}, which are located between $7.5$ and $8.2\ \mu m$, and $12.5$ and $13.5\ \mu m$. For a fixed observation range, a higher transmission rate occurs at a higher elevation angle due to thinner atmospheric layers in the LOS path compared to a lower elevation angle at the same wavelength. For a fixed observation angle, it is possible that a longer range results in a smaller transmission rate due to more atmospheric absorption. The residual plots indicate that the network is able to learn the transmission under different geometries almost equally because the error shows a variation dependent on wavelength but not on elevation angle or range. Additionally, this residual is at least one order magnitude smaller than the ground-truth transmission rate.

On the contrary, the upwelling radiance gets smaller when the elevation angle increases. This is because $L_{up}$ is a function of atmospheric emittance scattered or directly emitted to the LOS path integrated over the hemisphere. When the elevation angle is higher, the air is cooler and thinner for the same range of LOS path, resulting in less emitted energy received at the sensor. Additionally, its variation is more likely to be the reverse of the transmission. The reason to account for this is for any given material, the transmission, emissivity and reflectivity follow the equation: $\tau(\lambda) + \varepsilon(\lambda)+ r(\lambda) =1$. When the atmospheric transmission is high, its emissivity must be low, and vice versa.

The downwelling radiance is an independent variable on the target-sensor geometry which is the integration of atmospheric emittance over the hemisphere reaching at the target. Thus, the ground-truth downwelling radiance in \Fig{fig:downResidual} (a1) and (b1) at different angles and ranges are the same but only vary over wavelength. However, the network predicted downwelling radiance introduces a certain amount of errors, which are geometry dependent. Plots (a3) and (b3) show that the predicted downwelling radiance has a larger error at a lower elevation angle and longer range, especially for the wavelength bands between $7.5$ and $8.2\ \mu m$ where the atmospheric absorption is strong. 

BH data only has the observed at-sensor total radiance without any ground-truth atmospheric radiative component. To measure the model performance, the RMSE of the observed and derived at-sensor total radiance based on predicted $\widehat{L_{down}}$, $\widehat{L_{up}}$, and $\widehat{\tau(\lambda)}$ is reported in the following section, assuming a known target temperature and reflectivity. To validate the network on the unknown targets, the error of the retrieved target temperature and emissivity is presented in the \Sect{sec:retrieveEmiss}.

\subsection{BH At-sensor Total Radiance}
\begin{figure}
    \centering
    \includegraphics[width=\linewidth]{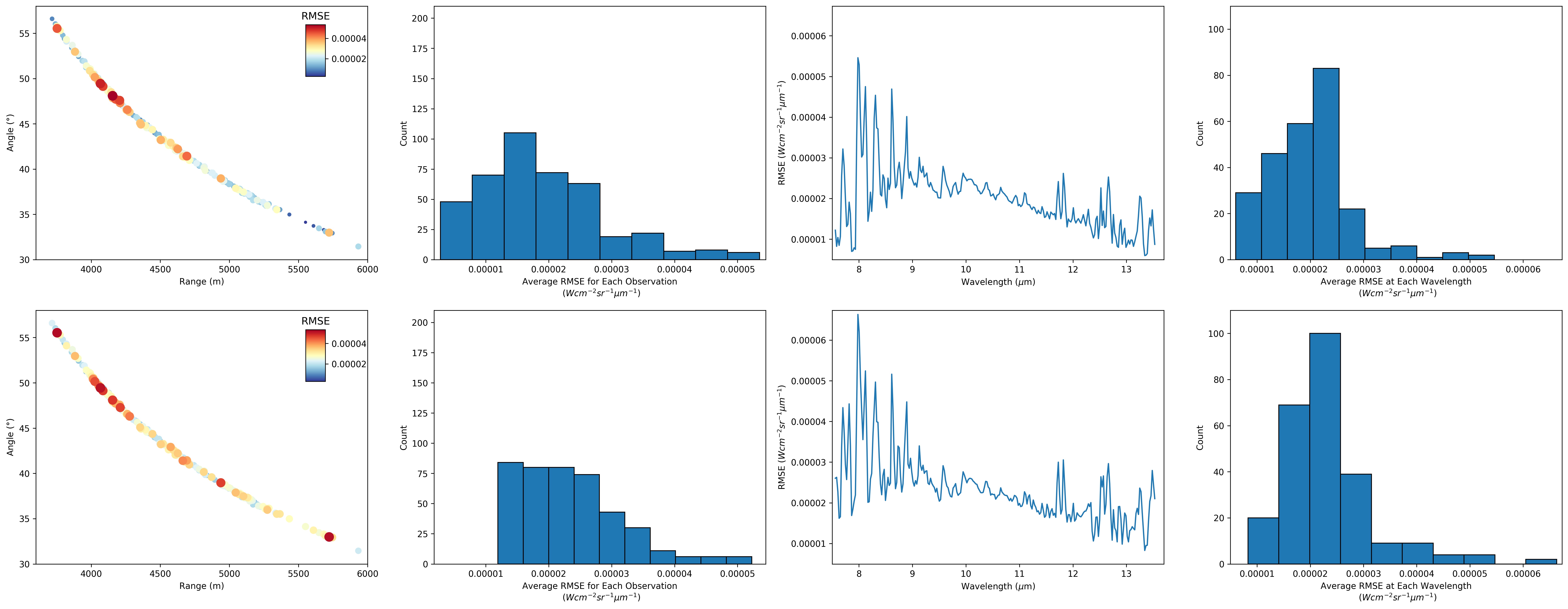}
    \caption{Comparison of two models, BH ill-posed training (First row), BH-MODTRAN mixed-training (Second row)}
    \label{fig:modelComp}
\end{figure}

\Fig{fig:modelComp} compares the ill-posed network trained only on BH data (first row) and the network trained on both MODTRAN and BH dataset (second row). Both trained models are evaluated on the BH test dataset to predict the atmospheric components (i.e., $\widehat{L_{down}}$, $\widehat{L_{up}}$, $\widehat{\tau(\lambda)}$) and calculate the at-sensor total radiance given the target true temperature and emissivity. The first and second columns are the average RMSE over all wavelengths under different geometries and its histogram. The last two columns show the average RMSE over all observations at different wavelengths and its histogram. 

From this figure, the average error of the derived at-sensor total radiance is smaller in the ill-posed model compared to the mixed-training network. However, this only indicates that the ill-posed network maintains the relationship of different radiative components towards the at-sensor total radiance in \Eq{equ:simpRTE} better than the mixed-training model but each component is still an ill-posed solution. The mixed-training model has introduced noise due to the intrinsic differences of the two datasets. In general, these two models are consistent with each other in the error distribution over observations and wavelengths. The influence of ill-posed predicted atmospheric radiative components needs to be evaluated on the retrieved target spectra to provide a fair comparison with the mixed-training model.

\subsection{Retrieved Target Temperature}\label{sec:retrievTemp}
To solve the temperature emissivity separation problem, one extra parameter is introduced, called $\epsilon$ for each material to represent its assumed emissivity. This $\epsilon$ is a constant, typically assumed as the mean value of true spectral emissivity over all 256 wavelength bands.

\subsubsection{MODTRAN Test Dataset}
\begin{figure}[ht]
	\centering
	\includegraphics[width=0.8\linewidth]{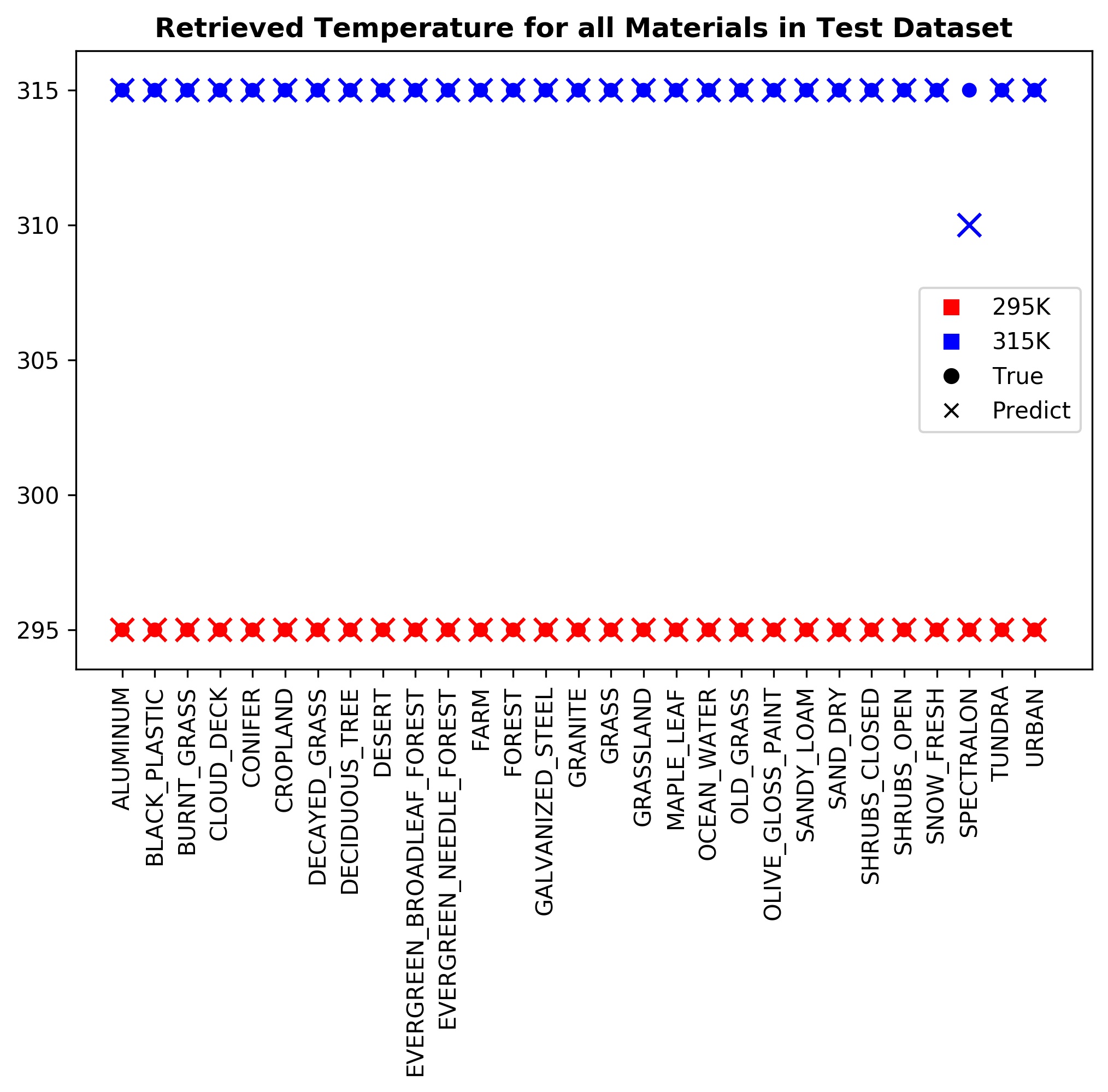}
	\caption{The retrieved target temperature for all 29 Materials in the testing dataset}
	\label{fig:retrieveT}
\end{figure}

To examine the accuracy of the grid search method given the assumed $\epsilon$, an experiment is conducted on 29 material at 2 different ground-truth temperatures (i.e., $295\ K$ and $315\ K$) under multi-geometries in the test dataset by using the proposed grid search method. \Fig{fig:retrieveT} shows the retrieved temperatures for 29 materials are the same as the ground-truth target temperatures, except for the spectralon material at $315\ K$, for which the estimated $T$ is $310\ K$.

\begin{figure}[ht!]
\begin{tabular}{c}
    \includegraphics[width=0.95\textwidth]{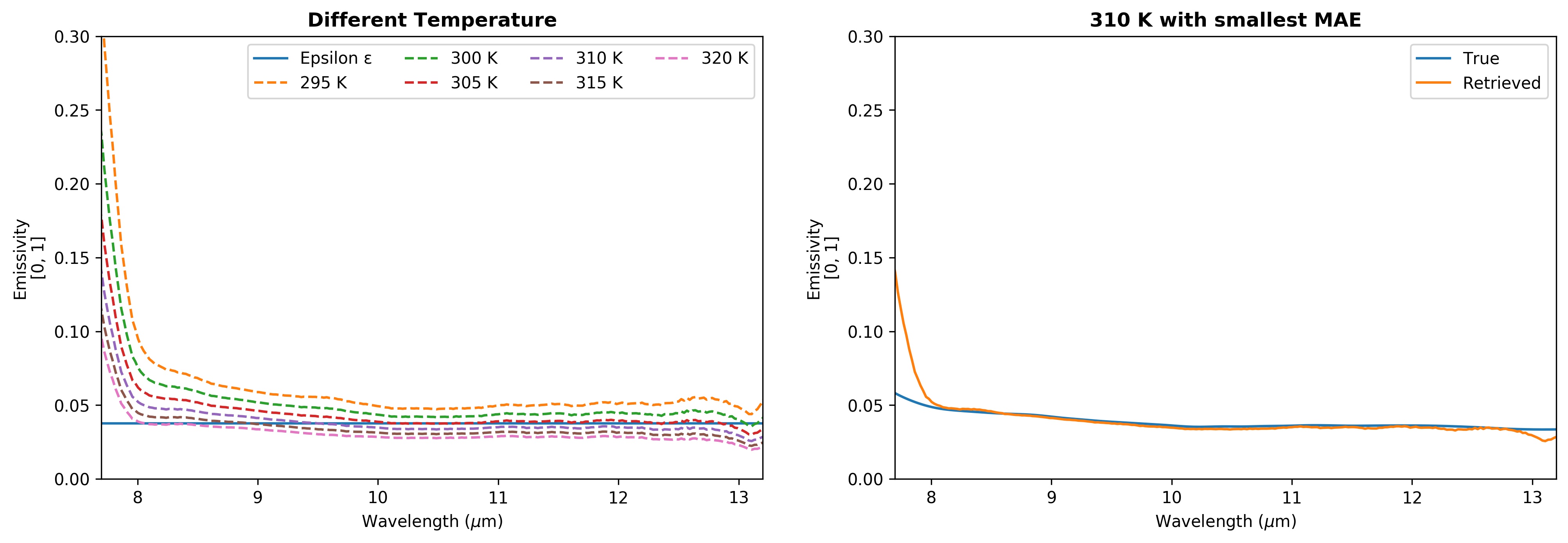}\\
    \kern 30pt (a) ~\kern 210pt ~ (b)\\
    \includegraphics[width=0.95\textwidth]{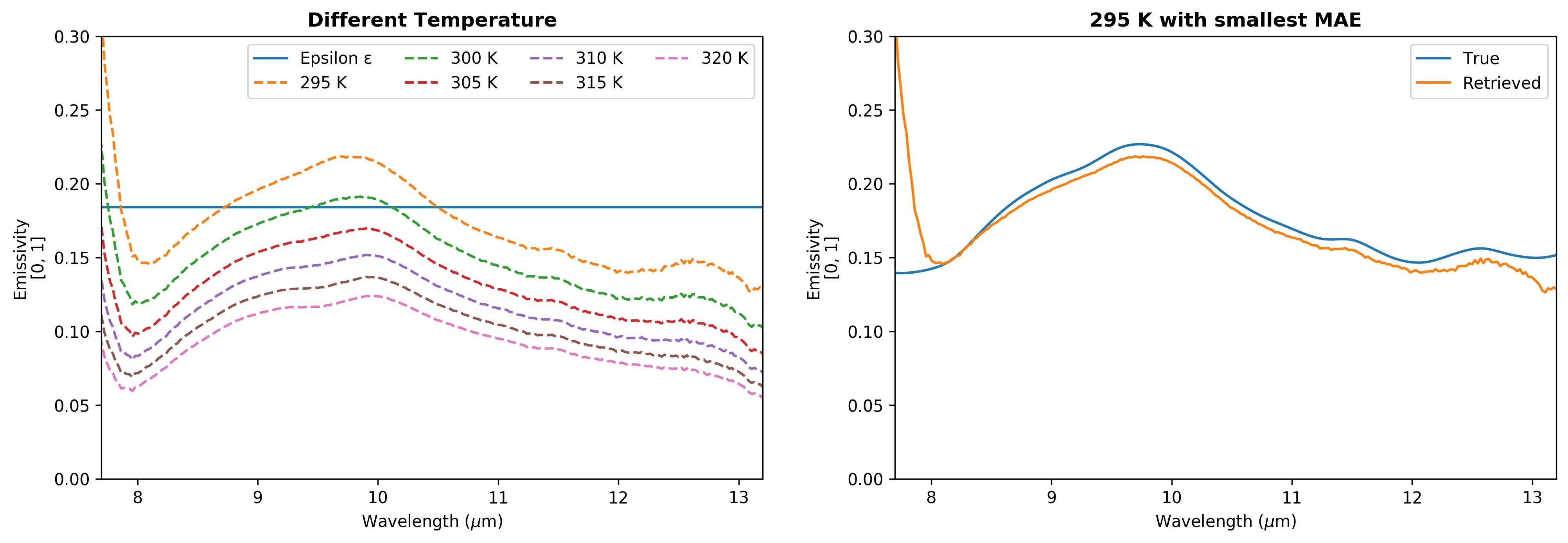}\\
    \kern 30pt (c) ~\kern 210pt ~ (d)\\
\end{tabular}
\caption{Retrieved emissivity for Spectralon (first row) and Aluminum (second row)}
\label{fig:tes}
\end{figure}

\Fig{fig:tes} (a) shows retrieved emissivity at different searched temperatures for spectralon with an assumed $\epsilon$ of 0.039. The retrieved emissivity at $310\ K$ and $315\ K$ are both close to the $\epsilon$ with a MAE of 0.004 and 0.0065 between $8\ \mu m$ and $13\ \mu m$, respectively. Plot (b) is the retrieved emissivity at estimated $310\ K$ versus the ground-truth spectralon emissivity. Even with a $5\ K$ temperature deviation from the true target temperature, the retrieved emissivity also successfully captures the spectralon's emissive spectra. 
Plots (c) and (d) show another example of the retrieved emissivity of Aluminum at different temperatures. The $\epsilon$ is set at 0.174 over 256 wavelength bands. The retrieved emissivity having the smallest MAE with this $\epsilon$ value is found at $295\ K$ which is also the ground-truth target temperature. The retrieved emissivity has a MAE of $0.00613$ with the ground-truth Aluminum emissivity.

Moreover, the values of retrieved emissivity are lower when the estimated temperature is higher. This is because $T$ has a positive relationship with $L_T$ in \Eq{equ:emissEq}. When $T$ increases, $\varepsilon(\lambda)$ decreases due to a larger denominator. When no extra information (e.g., $\epsilon$) could be provided to separate the temperature with emissivity, the grid search method could still give an estimation of the target spectral emissivity over 256 wavelengths. As \Fig{fig:tes} shows, all retrieved emissivities at different temperatures follow a similar pattern with the ground-truth emissivity at various scales.

\begin{figure}[!ht]
	\centering
	\includegraphics[width=\linewidth]{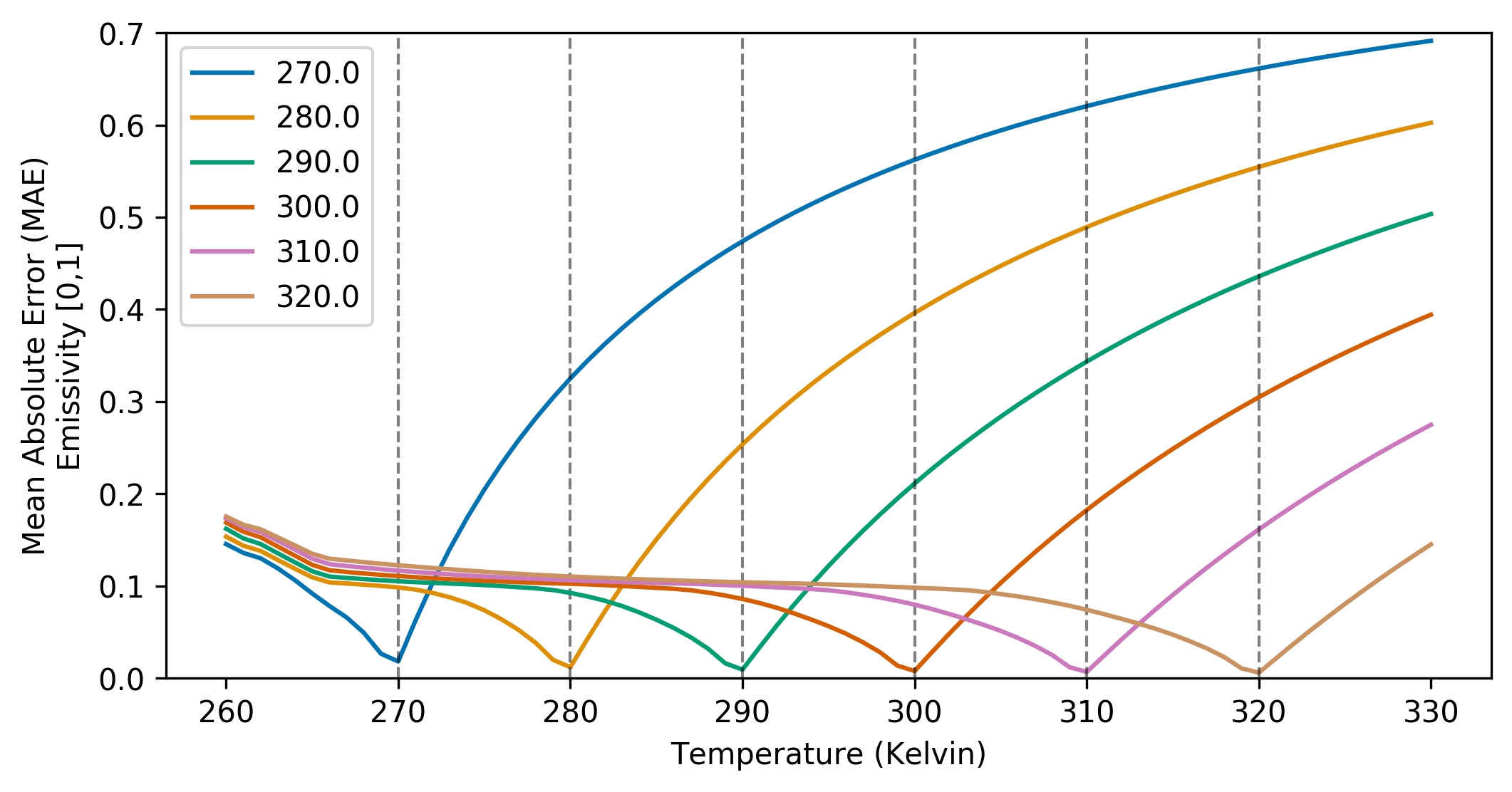}
	\caption{MAE of Retrieved emissivity when  estimated Temperature deviates from the Ground-truth Value}
	\label{fig:tempDeviation}
\end{figure}

To understand the influence of the deviation of the estimated temperature from the ground truth, the MAE of the retrieved emissivity estimated at different temperatures, is calculated for 29 materials under six different temperatures. \Fig{fig:tempDeviation} shows the MAE curve when the estimated temperature deviates from the ground truth and each color represents targets at a specific temperature. The retrieved emissivity is estimated at 71 different temperatures varying from $260\ K$ to $330\ K$ by every $1\ K$. Several observations are as follows:
\begin{enumerate}
    \item The error is smallest when the  estimated temperature is correct. When the deviation from the ground truth increases in either positive or negative direction, the error increases
    \item When the target is cooler, the MAE increases faster but when the target is hotter, the MAE increases slower as the deviation increases
    \item The underestimation is better than the overestimation for the temperature
\end{enumerate}

These MAE curves and observations from them may vary for different materials. For example, the retrieved emissivity of spectralon in \Fig{fig:tes} shows that the smallest MAE is not estimated at the correct target temperature. \Fig{fig:tempDeviation} represents the average of all materials' MAE curves. In general, a better accuracy of retrieved emissivity is achieved when the deviation of estimated temperature is smallest, however, within a range of $\pm 5\ K$, the MAE of retrieved emissivity is below 0.1. 

\subsubsection{BH Test Dataset}
The temperature of all pixels in the BH training dataset is set at $285\ K$ which is around the collected air temperature on April 18. This may not be exactly the same for all grass pixels but the under-estimation or an over-estimation of target temperature within $5\ K$ has been demonstrated to have a small error less than 0.1 shown in \Fig{fig:tempDeviation}. Moreover, some pixels extracted from the BH hyperspectral images in the test dataset may not be pure grass pixels. For different viewing angles, the pixel could be observed as a mixture of grass and soil background, thus its retrieved emissivity has a larger MAE with the ground truth. This means the retrieved emissivity is allowed to have a $\pm 0.1$ MAE error for pure grass pixels but the spectra over 256 wavelengths should be similar to the ground-truth target emissive spectra according to \Fig{fig:tes}. For the pixels mixed with a soil background, the MAE error could be larger when compared to the grass spectra. 

Taking these possible error sources into consideration, only the $MAE$ used in the grid search method for MODTRAN test dataset is not enough to support an accurate temperature search. Here, one more criterion $Norm\_MAE$ is introduced to identify at which estimated temperature, the retrieved emissivity has a more similar spectral curve with the ground-truth. At the temperature emissivity separation stage, the $Norm\_MAE$ is calculated as the MAE between the retrieved emissivity and the $\epsilon$ over all wavelengths and all observations, which both are normalized with their own respective minimum and  maximum values. 

\begin{figure}[ht!]
    \centering
    \includegraphics[width=\textwidth]{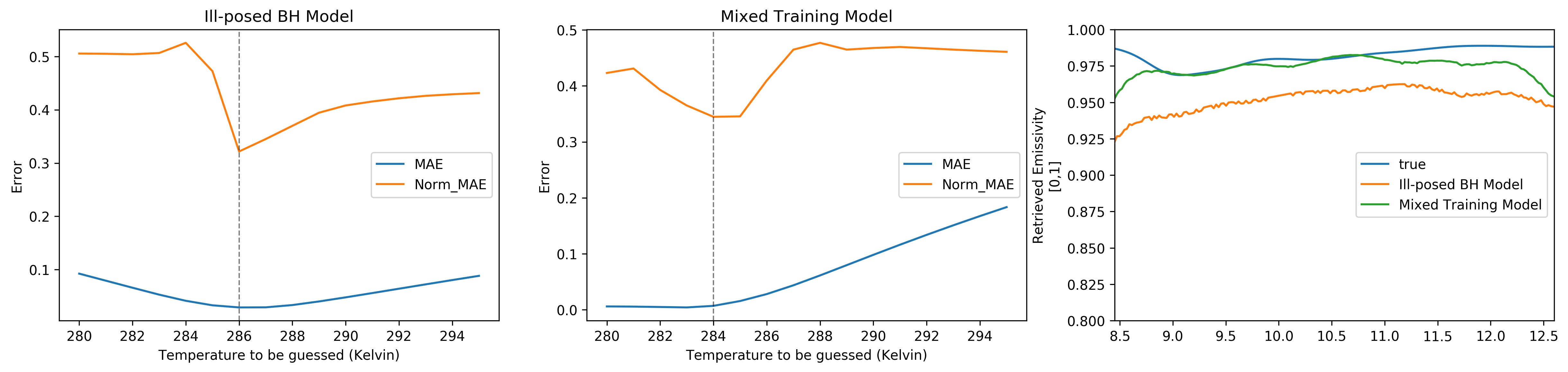}
    \caption{MAE, Norm\_MAE of different  estimated temperatures in the ill-posed BH model and mixed-training model as well as their retrieved emissivity at the  estimated temperature.}
    \label{fig:tempCriterial}
\end{figure}

The first two plots in \Fig{fig:tempCriterial} are values of $MAE$ and $Norm\_MAE$ at different temperatures in the ill-posed model and mixed-training model. Compared to the MODTRAN that chooses a temperature with the smallest $MAE$, the final estimated temperature in BH test dataset has the smallest $MAE+Norm\_MAE$. After adding the new criterion, the estimated target temperature in the mixed-training model changes from $283\ K$ to $284\ K$ while it is still the same ($286\ K$) in the ill-posed BH model. The last plot is the retrieved emissivity averaged over all observations in the test dataset at the estimated temperature. 

It is possible to see that the mixed-training model outperforms the ill-posed model to retrieve the target emissivity with a $5\%$ higher accuracy. The error gets larger towards the end after $12.5\ \mu m$ in the mixed-training model. To quantify the error caused by a mixture of the soil background, a pixel-wise analysis of the retrieved emissivity in the BH test dataset is done to identify observations that are outliers (mixed with the soil background).

\subsection{Retrieved Surface Emissivity}\label{sec:retrieveEmiss}
Given the  estimated temperature from the grid search method, all pixels' emissive spectra can be retrieved with their observed at-sensor total radiance $L_\lambda$ and predicted atmospheric components ($\widehat{L_{down}}$, $\widehat{L_{up}}$, $\widehat{\tau(\lambda)}$).

\subsubsection{MODTRAN Test Dataset}
\begin{figure}[ht!]
	\centering
	\includegraphics[width=\textwidth]{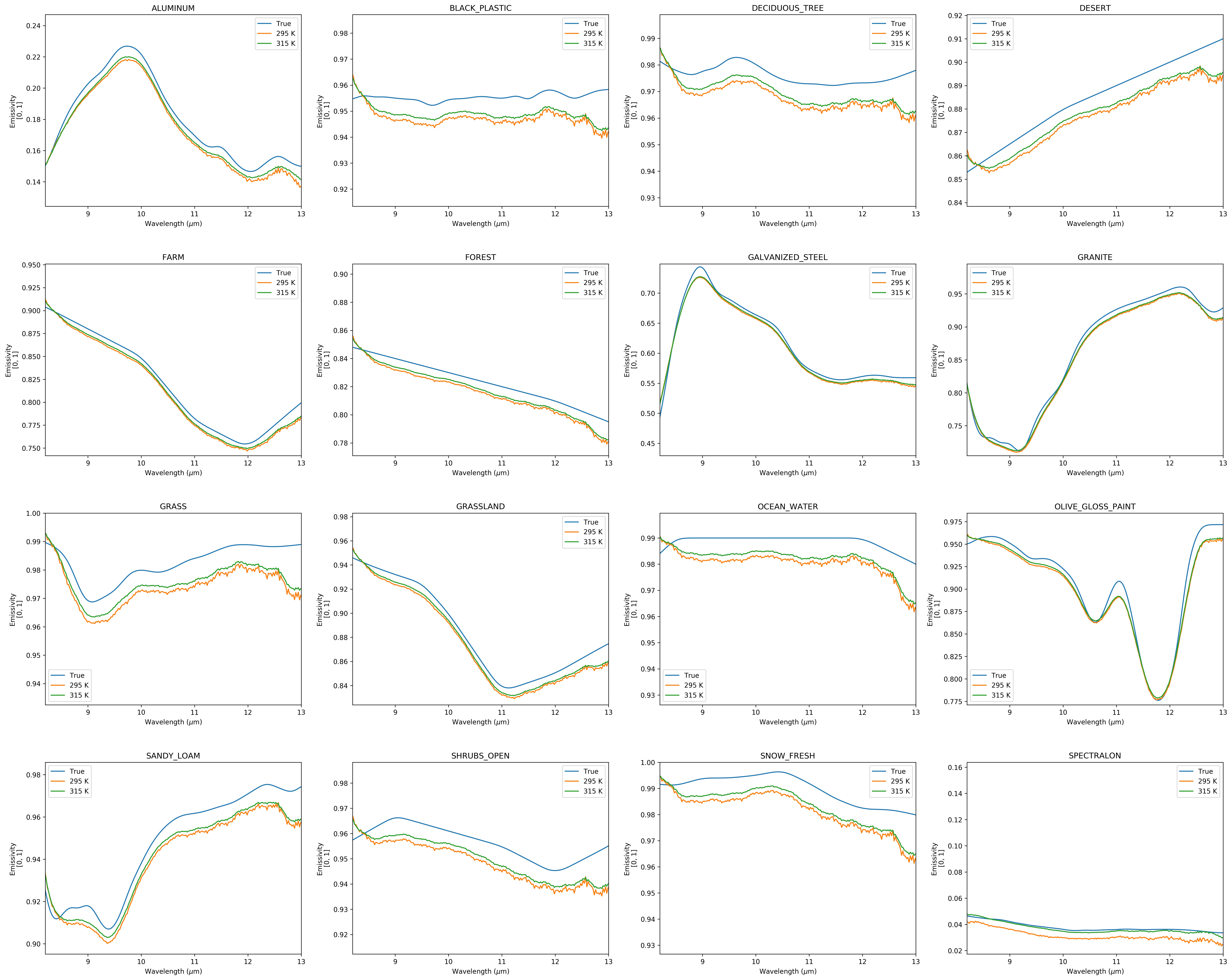}
	\caption{Retrieved emissivity for MODTRAN materials at  estimated Temperature}
	\label{fig:MODTRANemissRetri}
\end{figure}

\Fig{fig:MODTRANemissRetri} shows the retrieved emissivity for 16 materials whose emissive spectra are neither constant nor near constant (with a small perturbation) in the MODTRAN testing dataset between $8.5\ \mu m $ and $13\ \mu m$. In this test dataset, all materials are measured at two different temperatures (i.e., $295\ K$, $315\ K$). A grid search method is used to estimate the likely temperature, as shown in \Fig{fig:retrieveT} while the target emissivity is retrieved at the two estimated temperatures.  

The retrieved emissivity via the proposed hybrid network, has successfully captured the spectral properties of different materials. The network has a better performance on targets with smooth and continuously varying emissive spectra, such as aluminum, steel, granite, and olive gloss paint. There exist larger errors at peaks or dips of the spectra having dramatic changes within a small wavelength range or jagged spectral signatures, such as black plastic. 

The minimum and maximum MAE/RMSE of different materials over all wavelengths from $7.5\ \mu m$ to $13.5\ \mu m$ and observations in the test dataset are 0.0102/0.0137 and 0.0142/0.031, respectively. There exist larger errors at both two tails of the spectrum before $8.5\ \mu m$ and after $12.5\ \mu m$. This is because the atmospheric absorption around these wavelength bands is strong, which results in a low transmission for the radiance reaching at the sensor. Thus, small noise may cause larger errors of predicted transmission and downwelling radiance, shown in \Fig{fig:transResidual} and \Fig{fig:downResidual}. This is also indicated by that the maximum RMSE is two times larger than the maximum MAE because RMSE usually addresses larger errors with larger weights. The detailed investigation of the error source is given in the following section. 

\subsubsection{BH Test Dataset}
\begin{figure}[ht!]
    \centering
    \includegraphics[width=\textwidth]{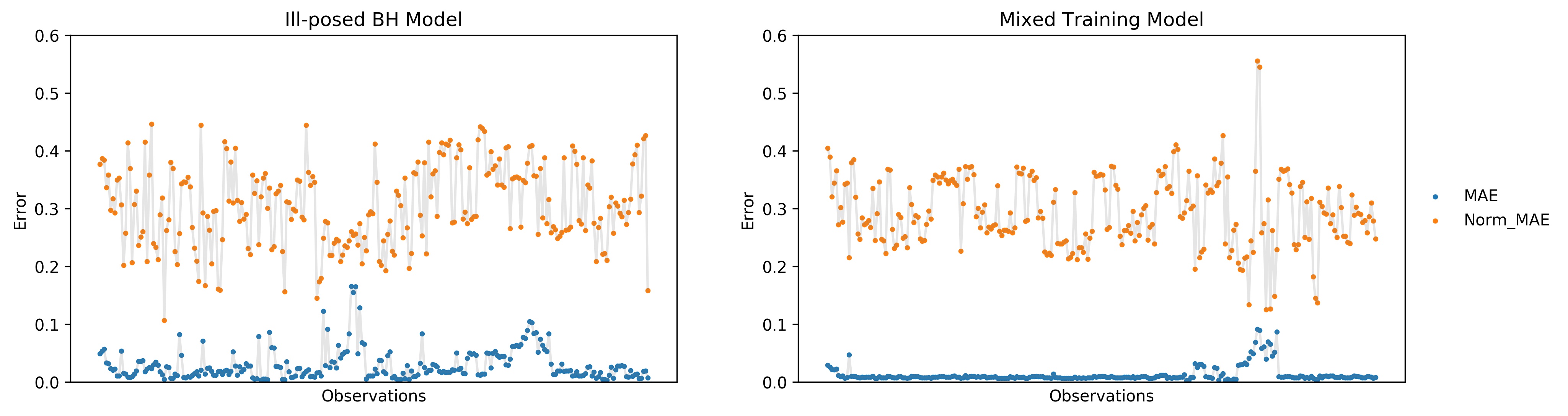}
    \caption{MAE and Norm\_MAE of different observations in the ill-posed BH model and mixed-training model}
    \label{fig:errorComp}
\end{figure}

\Fig{fig:errorComp} shows the $MAE$ and $Norm\_MAE$ for different observations at the estimated most-likely temperature. Here, the $Norm\_MAE$ is calculated as the MAE between the retrieved emissivity and the ground truth rather than $\epsilon$ over all wavelengths, which both are normalized with their own respective maximum and minimum values. Each pair of points in blue and orange represents the $MAE$ and $Norm\_MAE$ of a specific observation. 

The errors show that some pixels extracted from the BH images in the test dataset are not pure grass, thus their retrieved emissivities have a larger MAE error when compared to the ground truth. Both plots in \Fig{fig:errorComp} have a larger error at the same portion of observations towards the end of the spectrum. However, the error in the ill-posed model is more distributed over observations because it does not have enough constraints during the training process provided by the ground-truth atmospheric radiative components. Thus, the ill-posed model smooths out the error over all the observations via the network optimization and makes it harder to distinguish pure grass pixels from others. The mixed-training model forces the network to learn the global optimum with the guidance of the MODTRAN dataset. It can robustly separate the pure grass pixels from the pixels with a mixture of soil. $82.8\%$/$65.2\%$ of the observations in the mixed-training model could be categorized as pure grass pixels with a MAE under 0.02/0.01 while only $49.2\%$/$16.8\%$ of the observations in the ill-posed model are.

\begin{figure}[ht!]
    \centering
    \includegraphics[width=\textwidth]{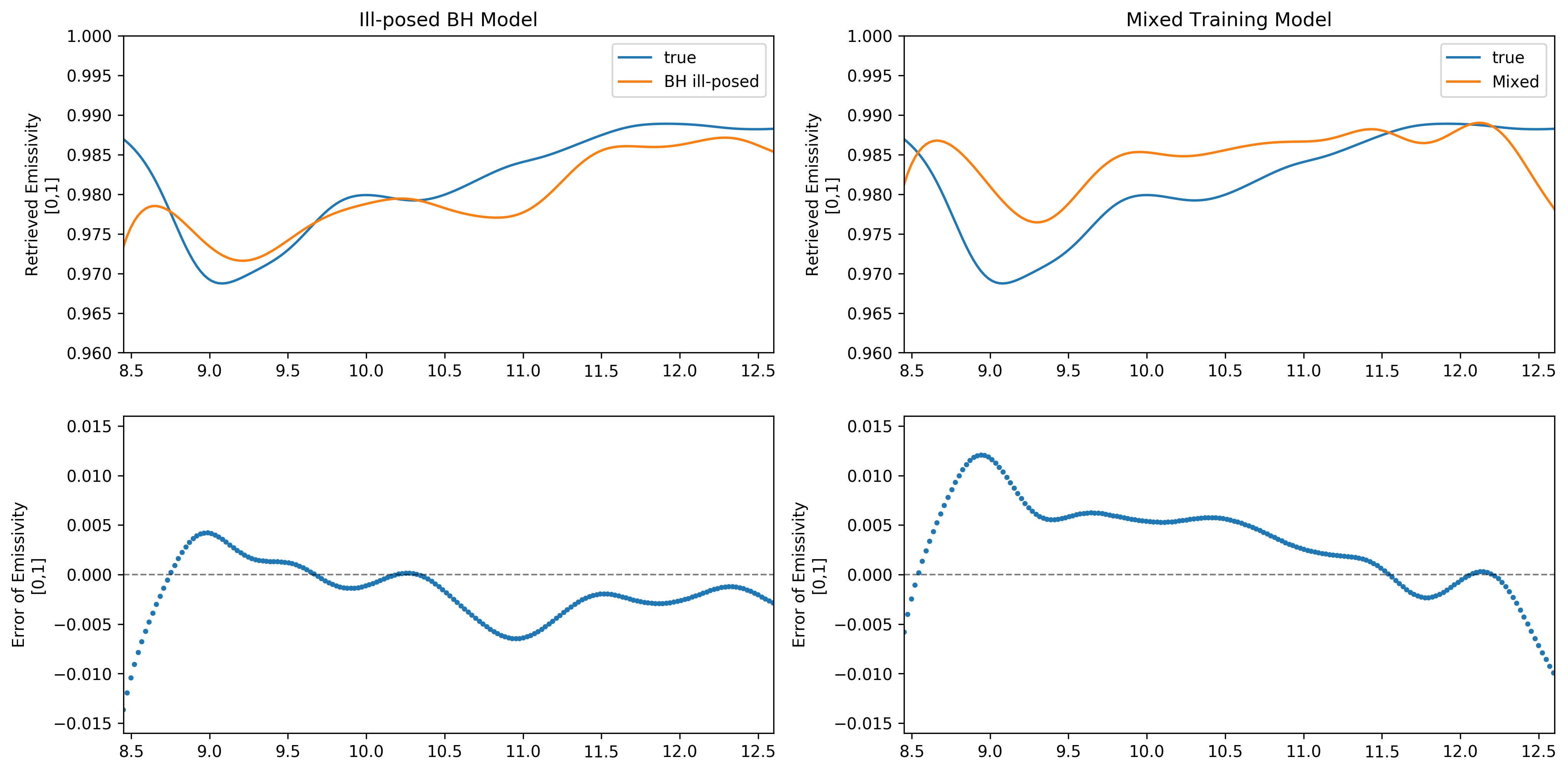}
    \caption{Retrieved emissivity for BH grass pixels at  estimated Temperature}
    \label{fig:BHemissRetri}
\end{figure}

\Fig{fig:BHemissRetri} shows the retrieved emissivity of pure grass pixels with MAEs under 0.01 in the BH test dataset. The ill-posed BH model underestimates while the mixed-training model overestimates the target emissivity. In general, the solution of the ill-posed model captures the rough emissive spectra but fails in the details. On the contrary, the mixed-training model shows a more similar emissive spectral curve with the ground truth but has a larger error, especially at the peaks or dips and at wavelengths longer than $12.2\ \mu m$. However, one point note is that more pixels in the mixed-training model have MAEs less than 0.01 compared to the ill-posed model ($65.2\%$ versus $16.8\%$ of total). Thus, the mixed-training model outperforms the ill-posed training model in the real-world application because it is more robust to distinguish pixels mixed with a background and can retrieve emissive spectra accurately compared to an ill-posed model.

\subsection{Error Investigation}
To explain why there is a larger error in both two tails of the retrieved emissivity from $7.5\ \mu m$ to $8.5\ \mu m$ and $13\ \mu m$ to $13.5\ \mu m$, several possible error sources are listed as below:
\begin{enumerate}
    \item Inaccurate assumptions of invariant atmosphere and pixels in open areas during data collection and data processing process.
    \item Intrinsic difference between MODTRAN simulated data and BH collected data. The middle-latitude spring-summer MODTRAN atmospheric model was used. However, despite this being the most similar model to the collection condition, it is important to remember that the aerosols, wind speed, air temperature, pressure, and humidity parameters are likely different from the real-world atmospheric profile during the BH data collection. It is then to be expected that simulations and observations will differ significantly because of the uncertainty associated with the atmospheric modeling.
    \item MODTRAN simulated data shows strong atmospheric absorption before $8.2\ \mu m$ and after $12.5\ \mu m$. This means the transmission value $\tau(\lambda)$ in equation \Eq{equ:emissEq} as denominator is small. Thus, any noise in the simulated or collected data or errors of network predictions in these wavelength bands is amplified on the retrieved emissivity.
\end{enumerate}

The solution for the first error source requires a great amount of real-world labeled training samples to train the network model. The second and third error sources are investigated by checking whether MODTRAN simulated data can satisfy \Eq{equ:simpRTE}.  For this study, MODTRAN data are simulated at the temperature of $295\ K$ for all 29 materials, and under different geometries as examples. 

The at-sensor target emittance $L_{emit}$ can be estimated after removing simulated atmospheric radiance reaching at the sensor. If \Eq{equ:simpRTE} is correct for the MODTRAN simulated data, $\frac{L_{emit}}{\tau(\lambda)}$ should be a constant for all observations of a specific target at a specific temperature of $T$, which is equivalent to the target surface-leaving emittance. However, estimating the target temperature with a known target spectral emissivity shows $\frac{L_{emit}}{\tau(\lambda)}$ is not constant, and vice versa.

\begin{figure}[ht]
	\centering
	\includegraphics[width=\textwidth]{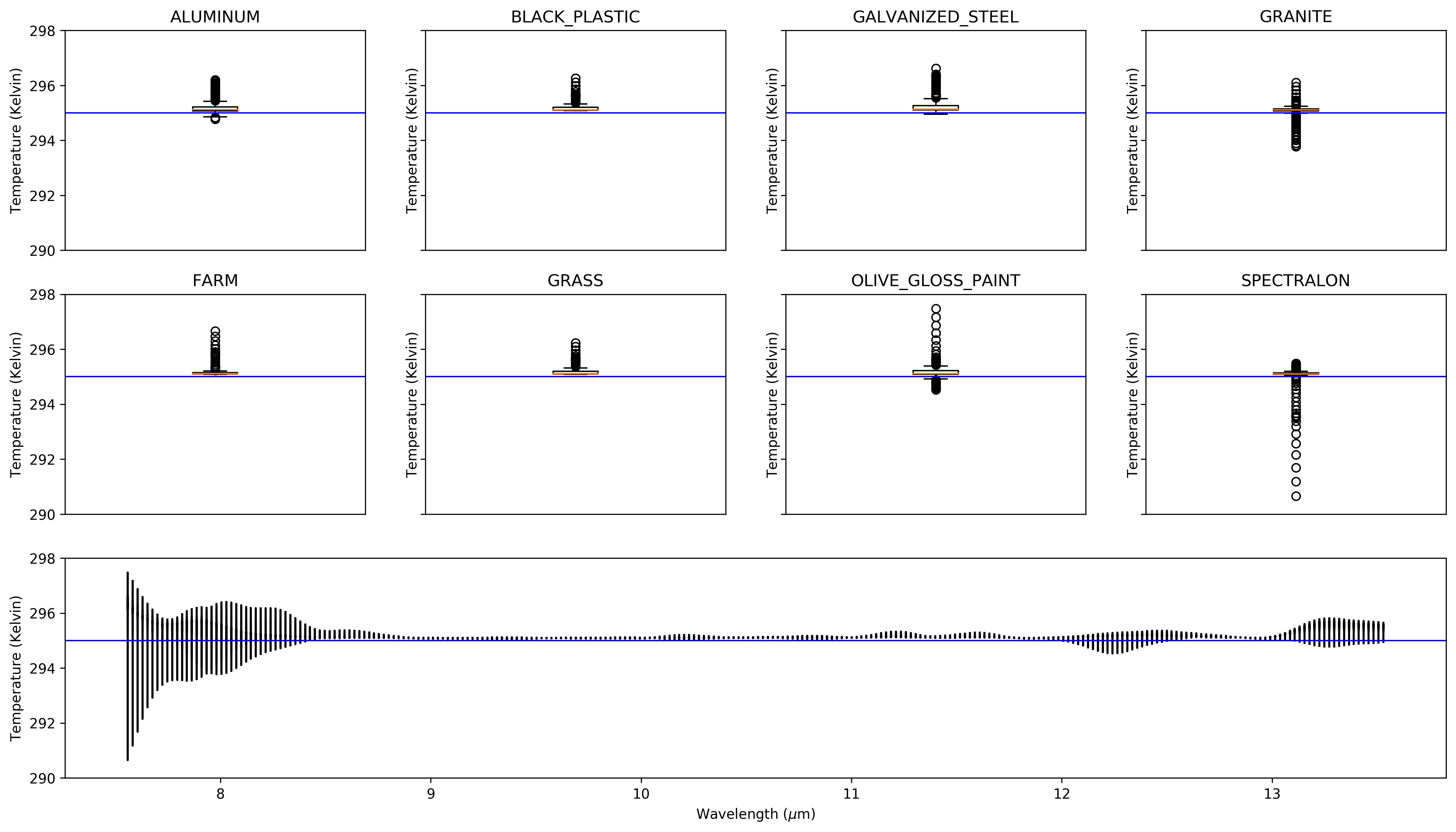}
	\caption{Equivalent Blackbody Temperature of MODTRAN Simulated Materials at 295 K}
	\label{fig:tempBoxplot}
\end{figure}

\Fig{fig:tempBoxplot} shows box-plots of eight typical materials with larger variances of retrieved temperatures over 256 wavelength bands. The temperature varies at different wavelength bands with a mean value around $295\ K$ and this variation is different for different materials. When this variation is projected into a barplot for all 29 materials at each wavelength band, there are larger errors between: 1) 7.5 and 8.5 $\mu m$, 2) 12.2 and 12.4 $\mu m$, and 3) 13.1 and 13.5 $\mu m$. This variation explains why the retrieved emissivity for each material has a larger deviation from the ground truth at two tails of wavelength bands when the temperature is fixed for all wavelengths. In other words, the temperature is a constant across all wavelengths. The spectral emissivity is also same for all observations of the same material. However, due to the intrinsic noise introduced by the MODTRAN radiative transfer model, there exists error during the temperature emissivity separation. This error varies over wavelengths and materials. The MAE of the simulated temperature of different materials over all wavelengths are within a range of $[0.173,\ 0.288]\ (K)$ while this MAE of different wavelengths over all materials is $[0.084,\ 1.382]\ (K)$. The error caused by different wavelengths is more significant than the error caused by different materials. 

The reason to account for this error is there exist three atmospheric strong absorption windows in the LWIR spectrum between $7.5$ and $8.5\ \mu m$, $12.2$ and $12.4\ \mu m$, as well as between $13.1$ and $13.5\ \mu m$. Water vapor has a strong absorption of the thermal infrared radiance between $5$ and $8.5\ \mu m$ while carbon dioxide is another strong absorber between $12$ and $13\ \mu m$. The target temperature is indeed simulated at $295\ K$ over all wavelengths but small noise is introduced by the MODTRAN radiative transfer model within these strong atmospheric absorption bands. When this simulated at-sensor total radiance is used as ground-truth, this error is transferred to the target emissivity estimation. This intrinsic problem is common in many research topics when the atmosphere attenuates the radiance greatly and makes the target emissive spectra impossible to retrieve. 

\section{Conclusion}
This research proposes a novel hybrid neural network to characterize the atmosphere and retrieve surface emissivity without additional meteorological data, using multi-scan hyperspectral images collected from different geometries. The design of this network has fully considered the realistic noise as well as the missing labeled data, where the network is mainly convolutional-based but involved with different fully connected layers to either process geometric information or generate geometric-independent predictions. The novelty of this research is that it implements the causal relationship of different geometric factors for each radiative component into the network structure by breaking down the learning process into multiple sub-tasks, which can be processed parallelly while their physical connections are still maintained. 

To solve the ill-posed atmospheric characterization during the training process, the proposed solution exploits both real-world collected BH hyperspectral images with only the observed at-sensor total radiance, as well as the simulated MODTRAN data with ground-truth atmospheric components. The loss functions used are required to: 1) guide the network to converge towards a plausible solution based on simulated data; and 2) update the network parameters to find the real local optimum using observed real-world BH data. With a small amount of pixels of known grass pixels extracted from the BH images, the proposed network is able to provide enough constraints to the network to convert the at-sensor total radiance to the real surface emissive spectra.

The output of this network consists of the three atmospheric radiative components, namely downwelling, upwelling and atmospheric transmission, which are required to perform the correction. The at-sensor total radiance and target emittance reaching at the sensor can be further calculated given the predicted results, the surface temperature and emissivity. The stability and the accuracy of predicted atmospheric components are tested giving the wrong target temperature and emissivity. A target temperature and emissivity separation method is proposed to estimate the target temperature. Given the observed at-sensor total radiance, the proposed method estimates the atmospheric components to correct for the atmosphere, and once the target temperature is also estimated, the surface emissive spectra can be retrieved.

This work only focuses on the LWIR spectrum, where the solar contribution is negligible compared to the thermal signatures. However, the proposed geometry-dependent network is expected to work for all parts of the spectrum, although different assumptions and simplifications must be taken. The current work was tested only using grass pixels. In future work, a comprehensive analysis of BH collected hyperspectral images should be conducted to better understand the accuracy of the retrieved spectral signatures for different ground surfaces.

\section*{Acknowledgment}
This research was supported by the Defense Advanced Research Projects Agency (DARPA) award FA-8650-19-1-7905.

\bibliography{mybib}
\end{document}